\documentclass[aps,preprint,showpacs,floatfix]{revtex4-1}
\usepackage{bm}
\usepackage{amsmath}
\usepackage{epsfig}
\usepackage{rotating}
\usepackage{graphicx}

\def\pr{\prime}
\def\be{\begin{equation}}
\def\lan{\left\langle}
\def\ran{\right\rangle}
\def\ee{\end{equation}}
\def\barr{\begin{array}}
\def\earr{\end{array}}

\def\l{\left}
\def\r{\right}
\def\dis{\displaystyle}
\def\ed{\end{document}}
\def\f{\frac}

\def\cs{{\bf S}}

\def\eh{{\hat{E}}}

\def\sh{\hat{\Sigma}}

\def\on{\overline{n}}
\def\ed{\end{document}}
\begin{document}

\title{Bivariate moments of the two-point correlation function for embedded Gaussian unitary ensemble with $k$-body interactions}

\author{V. K. B. Kota\footnote{vkbkota@prl.res.in}}
\affiliation{Physical Research  Laboratory, Ahmedabad 380 009, India}

\begin{abstract}

Embedded random matrix ensembles with $k$-body interactions are well established
to be appropriate for many quantum systems. For these ensemble the two point
correlation function is not yet derived though these ensembles are introduced 50
years back. Two-point correlation function in eigenvalues of a random matrix
ensemble is the ensemble average of the product of the density of eigenvalues at
two eigenvalues say $E$ and $E^\pr$. Fluctuation measures such as the number
variance and Dyson-Mehta $\Delta_3$ statistic  are defined by the two-point
function and so also the variance of the level motion in the ensemble. Recently,
it is recognized that for the embedded ensembles with $k$-body interactions the
one-point function (ensemble averaged density of eigenvalues) follows  the so
called $q$-normal distribution. With this, the eigenvalue density can be
expanded by starting with the $q$-normal form and using the associated
$q$-Hermite polynomials $He_\zeta(x|q)$. Covariances $\overline{S_\zeta
S_{\zeta^\pr}}$ (overline representing ensemble average) of the expansion
coefficients $S_\zeta$ with $\zeta \ge 1$ here determine the two-point function
as they are a linear combination of the bivariate moments $\Sigma_{PQ}$ of the
two-point function. Besides describing all these, in this paper derived are
formulas for the bivariate moments $\Sigma_{PQ}$ with $P+Q \le 8$, of the
two-point correlation function, for the embedded Gaussian unitary ensembles with
$k$-body interactions [EGUE($k$)] as appropriate for systems with $m$ fermions
in $N$ single particle states. Used for obtaining the formulas is the $SU(N)$
Wigner-Racah algebra. These formulas with finite $N$ corrections are used to
derive formulas for the covariances $\overline{S_\zeta S_{\zeta^\pr}}$ in the
asymptotic limit. These show that the present work extends to all $k$ values the
results known in the past in the two extreme limits with $k/m \rightarrow 0$
(same as $q \rightarrow 1$) and $k=m$ (same as $q=0$). 
\end{abstract}


\maketitle

\section{Introduction}

Classical random matrix ensembles, i.e. the Gaussian orthogonal, unitary and symplectic ensembles (GOE, GUE and GSE) are well known now in physics and need no introduction \cite{Porter,Mehta,RMT-book}. Hamiltonians ($H$) for atoms, atomic nuclei, molecules, mesoscopic systems such as quantum dots etc. consist of a mean-field one-body part and a residual two-body interaction. With the two-body part sufficiently strong, energy levels of these systems in general exhibit quantum chaos and the appropriate random matrix ensembles for describing this, as recognized first in nuclear shell model studies \cite{PLB1,PLB2,MF,Br-81}, are the so-called embedded ensembles (EE) generated by $k$-body interactions [EE($k$)] in many-particle ($m$-particle with $m >k$) spaces (assumed is that the particles are in $N$ number of single particle states with $N >> m$). In particular, the embedded Gaussian orthogonal and unitary ensembles generated by $k$-body interactions [EGOE(k) and 
EGUE(k)], applicable to many fermion systems, have received considerable attention in the last two decades. Remarkably, for $m >> k$ (with $N >> m$), these ensembles generate Gaussian eigenvalue densities, i.e. the one point function in the eigenvalues (one, two and higher point functions are defined by Dyson \cite{Dyson-1}). Here, it is important to note that for $m=k$, EE will reduce to the classical ensembles 
giving the well known Wigner semi-circle form for the one-point function \cite{MF,Br-81,BRW}. This important result is seen in large number of numerical calculations and it is also proved analytically \cite{MF,BRW,Ko-05,SM}. With $E$ denoting eigenvalues and $\rho(E)$ the eigenvalue density for a given member of an ensemble of random matrices, the one point 
function is $\overline{\rho(E)}$ where the overline indicates ensemble average.

Turning to the two-point correlation function, though large number of EGOE calculations showed that the spacing distribution, number variance and the Dyson-Mehta $\Delta_3$ statistic \cite{DyMe} and other measures of level fluctuations follow GOE, till today there is no success in deriving the two-point correlation function $\overline{\rho(x) \rho(y)}$ for EGOE(k) or EGUE(k) even in the limit of $k <<m$. Earliest attempt is due to French \cite{MF,Br-81,JBF-M} who has shown that EGOE($k$) in the dilute limit (with $k$ finite, $N \rightarrow \infty$, $m \rightarrow \infty$ and $m/N \rightarrow 0$) generates average-fluctuation separation that is absent in classical Gaussian ensembles. However, experimental confirmation of this feature is not yet available nor the formula for the two point function.  Next attempt is due to Verbaarschot and Zirnbauer \cite{VeZi}. This is followed by an attempt due to Weidenm\"{u}ller and collaborators \cite{BRW,BW-1}. However, as shown  by Srednicki later \cite{Serd}, the results in \cite{BRW} for the nature of level fluctuations generated by EGUE($k$) are inconclusive. A significant result due to Weidenm\"{u}ller et al is that EE generate the so called cross correlations that are absent in classical ensembles;
see \cite{PW,Ko-book} for results regarding cross correlations in EE. Here also, definitive experimental tests of these are not yet available.

Recently, a new direction in exploration of EE has opened up
with the analysis of quantum chaos in the Sachdev-Ye-Kitaev (SYK) model using
random matrix theory by Verbaarschot and collaborators
\cite{Verb1,Verb2,Verb3,Verb4,Verb5,Verb6}. Most significant result in these papers, for the present purpose, is the recognition that the so-called $q$-normal distribution indeed gives the eigenvalue density in the SYK model. Bryc, Szablowski, Ismail and others \cite{Boze,Bryc, Sza-1,Sza-2,Ismail} earlier clearly showed that this $q$-normal distribution (see Section \ref{sec2} for definition and other mathematical details) has a purely commutative and classical probabilistic meaning. With the $q$-normal reducing to Gaussian form for $q=1$ and semi-circle form for $q=0$, immediately shows that EE($k$) will generate $q$-normal form for the eigenvalue densities. Remarkably, it is seen that the lower order moments (up to 8th order) of the eigenvalue density (one-point function) generated by EE($k$) are essentially identical to the lower order moments given by $q$-normal distribution \cite{qMK-1,qMK-2} with the fourth moment (this depends on $k$) determining the value of the $q$ parameter. With this, there is a possibility that expansions for $\rho(E)$ starting from the $q$-normal form using the associated $q$-Hermite polynomials may allow us to understand
the two-point function for EE($k$) with $k$ changing from $k=2$ to $m$ ($k=1$ appears to be special \cite{One,Pandey}) just as it was done in the past for the classical Gaussian ensembles and also adopted for EGOE($k$) with $k << m$ \cite{MF,FMP,Br-81}.
Interestingly, expansion involving $q$-Hermite polynomials is
also employed in investigating level fluctuations in the SYK model \cite{Verb4}. Following this, we have revisited the problem of deriving the two-point correlation function for EE($k$) and analytical formulas for the bivariate moments (to order 8) of the two-point function for EGUE($k$) are presented in this paper. These will determine the covariances of the expansion coefficients appearing in the $q$-Hermite polynomial expansion of the eigenvalue density of the EGUE($k$) ensemble members. It is expected that these results may yield the two-point function for EGUE($k$) in the near future. Now we will give a preview.  

In \ref{sec2}, firstly for completeness, EGUE($k$) is defined. Secondly,
introduced are the two-point function and its integral version along with their
relation to the number variance, $\Delta_3$ statistic and variance of the level
motion in the ensemble. In addition, the $q$-normal form $f_{qN}$ along with
$q$-Hermite polynomials are defined and collected are some of their properties.
In Section \ref{sec3}, using the expansion of the eigenvalue density in terms of
$q$-normal $f_{qN}$ and $q$-Hermite polynomials $He_\zeta(x|q)$, it is shown that the covariances of the expansion coefficients $S_\zeta$ ($\zeta=1$, $2$, \ldots, $\infty$) are related in a simple manner to the bivariate moments $\Sigma_{PQ}$ of the two-point function. Following this, in Section \ref{sec4} derived are formulas for the bivariate moments $\Sigma_{PQ}$ of the two point function for EGUE($k$) for $P+Q \le 8$ using the formulation in terms of $SU(N)$ Wigner-Racah algebra as described in \cite{Ko-05}. In Section \ref{sec5} presented are asymptotic limit formulas for the covariances $\overline{S_\zeta S_{\zeta^\pr}}$ for EGUE($k$) ensemble. In addition, some general structures indicated by these formulas are also discussed and an expansion
for the number variance is given. Finally, Section \ref{sec6} gives conclusions.

\section{Preliminaries : EGUE($k$), Two-point function, $q$-normal distribution, $q$-Hermite polynomials}
\label{sec2}

\subsection{EGUE($k$) definition}

Given a system of $m$ spinless fermions distributed in $N$ degenerate single particle (sp) states and interacting via $k$-body $(1 \leq k \leq m)$ interactions, the EGUE($k$) in $m$ fermions spaces is generated by representing the $k$-particle $H$ by GUE. For a more precise definition, firstly consider the sp states (denoted by $\nu_i$) in increasing order, $\nu_1 \leq \nu_2 \leq \cdots \leq \nu_N$. Now, a random $k$-body $H$ in second quantized form is,
\be
H(k) = \dis\sum_{\alpha,\;\beta} \; V_{\alpha,\beta}(k) \; \psi^\dagger(k; \alpha) \; \psi(k;\beta) \;.
\label{eqa-1}
\ee 
Here, $\alpha$ (similarly $\beta$) are $k$-particle states (configurations) $\l.\l|\nu^o_1,\nu^o_2,\ldots,\nu^o_k\r.\ran$ in occupation number representation; $\nu^0_i$ are occupied sp states. Distributing $k$ fermions (following Pauli's exclusion principle) in $N$ sp states will generate complete set of these distinct configurations ($\alpha$, $\beta$, $\ldots$) and total number of these configurations is $\binom{N}{k}$. Operators $\psi^\dagger(k; \alpha)$ and $\psi(k;\beta)$ respectively are $k$-particle creation and annihilation operators, i.e. $\psi^\dagger(k; \alpha) = \prod_{i=1}^{k} a^\dagger_{\nu^{\alpha}_i}$ and $\psi(k;\beta) = \prod_{j=1}^{k} a_{\nu^{\beta}_j}$; here for example $\nu_i^{\alpha}$ is $i$-th occupied sp state for the $k$-particle configuration $\alpha$.
The one-particle creation ($a^\dagger_{\nu_i}$) and annihilation ($a_{\nu_j}$) operators obey the usual anti-commutation relations. In Eq. (\ref{eqa-1}), $V_{\alpha,\beta}(k)$ matrix is chosen to be a $\binom{N}{k}$ dimensional GUE in $k$-particle spaces ($V$ matrix is complex hermitian). That means $V_{\alpha,\;\beta}(k)$ are anti-symmetrized $k$-particle matrix elements chosen to be randomly distributed independent Gaussian variables with zero mean and variance
\be
\overline{V_{\alpha,\beta}(k) \; V_{\alpha^\prime,\beta^\prime}(k)} = v^2 \; \delta_{\alpha,\beta^\prime} \;\delta_{
\alpha^\prime,\beta} \;.
\label{eqa-2}
\ee
Here, the bar denotes ensemble averaging and we choose $v=1$ without loss of generality. Distributing the $m$ fermions in all possible ways in the $N$ states generates the many-particle basis states (configurations) $\l.\l|\nu^o_1,\nu^o_2,\ldots,\nu^o_m\r.\ran$ in occupation number representation defining a $\binom{N}{m}$ dimensional Hilbert space. Action of the Hamiltonian operator $H(k)$ defined by Eq. (\ref{eqa-1}) on the above many-particle basis states generates a $H$ matrix ensemble in $m$-particle spaces with dimension $\binom{N}{m}$ and this is the EGUE$(k)$ ensemble - it is a random matrix ensemble in $m$-particle spaces generated by $k$-body interactions. Note that EGUE($k$) has three parameters $(N,m,k)$. See \cite{Br-81,BW-1,Ko-book,Small} for further details regarding not only EGUE($k$) but also for EGOE($k$), EGSE($k$) and many other extensions of embedded ensembles including those for interacting boson systems. In the present paper we restrict to EGUE($k$).

\subsection{Two-point function}

Let us begin with the ensemble averaged eigenvalue density or the one-point function $\overline{\rho(E)}$
of EGUE($k$)
where $\rho(E)$ is the eigenvalue density (normalized to unity and usually it is called frequency function in statistics) for each member of EGUE($k$); $E$ denotes energy eigenvalues and some times we will use $x$ or $y$ to denote eigenvalues. Integral version of $\rho(E)$ is the distribution function $F(x)$ (also called stair case function),
\be
F(x) = d\,\dis\int_{-\infty}^x \rho(E)\, dE\;.
\label{fl-1}
\ee
Note that $F(x)$ gives number of levels up to the eigenvalue
$x$ and $d$ is total number of eigenvalues, i.e. dimension of the given EGUE($k$). Now, the two-point correlation function $S^{\rho}(x,y)$ for the eigenvalues and its integral version $S^F(x,y)$ are (here and elsewhere in this paper mostly we employ the notations used in \cite{Br-81}),
\be
\barr{rcl}
S^{\rho}(x,y) & = & \overline{\rho(x)\;\rho(y)} -{\overline{\rho(x)}}\;{\overline{\rho(y)}}\;,\\
S^F(x,y) & = & d^2\; \dis\int_{-\infty}^x \dis\int_{-\infty}^{y} S^\rho(x^\pr ,y^\pr) dx^\pr dy^\pr \;\;=\;\;\overline{F(x)\;F(y)} -{\overline{F(x)}}\;{\overline{F(y)}}\;.
\earr \label{fl-2}
\ee
From Eq. (\ref{fl-2}), as bar denotes ensemble average, it is clear that $S^\rho$ (and $S^F$) gives
measures for level fluctuations and the simplest two-point measure is the number variance $\Sigma^2(\on)$. Say, there are $n$ number of levels between energies $x$ and $y$. Then $n=F(x)-F(y)$ and similarly $\on = \overline{F(x)} - \overline{F(y)}$.
With this, a measure for fluctuation in number of levels, with $\on$ the average number of levels, is the number variance 
$\Sigma^2(\on) = \overline{(n-\on)^2}$ and this is simply related to $S^F(x,y)$, 
\be
\Sigma^2(\on) = S^F(x,x) + S^F(y,y) -2 S^F(x,y)\;.
\label{fl-3}
\ee
In addition, the $\Delta_3$ statistic is simply related to $\Sigma^2(\on)$ \cite{Br-81},
\be
\Delta_3(\on) = \dis\frac{2}{\overline{n}^4} 
\dis\int_0^{\overline{n}}\,\l(
\overline{n}^3 - 2\overline{n}^2r + r^3\r)\,\Sigma^2(r)\,dr \;.
\label{fl-4}
\ee
Further, an approach to study $S^F(x,x)$ is to examine level motion in the
ensemble. For example, variance of the fluctuation in a eigenvalue $E$, measured in units of the local level spacing $\overline{D(E)}$ is denoted by
$\overline{\delta E^2}/\overline{D(E)}^2$. This is often called level motion variance. Then, it is easy to see that the variance of level motion is
\be
\overline{\delta E^2}/\overline{D(E)}^2 = S^F(E,E)\;.
\label{fl-4a}
\ee
Similarly, $S^F(x,y)$ and $S^\rho(x,y)$ can be probed or constructed using the bivariate moments ${\widetilde{\Sigma}}_{PQ}$ of $S^{\rho}(x,y)$,
\be
{\widetilde{\Sigma}}_{PQ} = \dis\int x^P\, y^Q \,S^{\rho}(x,y)
\,dx dy = \overline{\lan H^P\ran \lan H^Q\ran} -\overline{\lan H^P\ran}\;\;\overline{\lan H^Q\ran}\;.   
\label{fl-5}
\ee
With $\l.\l| \alpha_i\r.\ran$, $i=1,2,\ldots,d$ denoting the $m$ fermion basis states, the $P$-th moment of $\rho(E)$ is $\lan H^P\ran = d^{-1} \, tr(H^P)$ where $tr(H^P)$ is the trace of $H^P$ in $m$ fermion space. Note that $tr(H^P)= \sum_i \lan \alpha_i \mid H^P \mid \alpha_i\ran = \sum_i (E_i)^P$ as traces are invariant under a unitary transformation (also, $\sum_i (E_i)^P = d\,\int E^P \rho(E)\, dE$). It is easy to see from Eq. (\ref{fl-5})
that ${\widetilde{\Sigma}}_{PQ}={\widetilde{\Sigma}}_{QP}$ and
${\widetilde{\Sigma}}_{P0}=0$. Also, with
\be
\Sigma_{PQ} = \overline{\lan H^P\ran \lan H^Q\ran}\;,
\label{fl-51a}
\ee
we have $\Sigma_{P,0} = \overline{\lan H^p\ran^m}$, the $P$-th moment of $\overline{\rho(E)}$. Our purpose in this paper is to derive formulas for the bivariate moments $\Sigma_{PQ}$ with $P+Q \le 8$ (these are given in Section \ref{sec4}) as they will determine the lower order terms in an expansion of the two-point function and this is discussed in more detail in Section \ref{sec3}. Before turning to these, in the next subsection introduced are the $q$-normal distribution and $q$-Hermite polynomials as the eigenvalue density for EE($k$) (well demonstrated for EGOE($k$) and EGUE($k$) in \cite{qMK-1}) is close to $q$-normal and this reduces to Gaussian form for $k << m$ and semi-circle for $k=m$. Thus, the $q$ normal form covers all $k$ values.

\subsection{$q$-normal distribution and $q$-Hermite polynomials}

Firstly, $q$ numbers $[n]_q$ are defined by (with $[0]_q=0$)
\be
\l[n\r]_q = \dis\frac{1-q^n}{1-q} = 1+q + q^2 + \ldots+q^{n-1}\;.
\label{fl-6}
\ee
Note that $[n]_{q \rightarrow1}=n$. Similarly, $q$-factorial $[n]_q! = \dis\Pi^{n}_{j=1} \,[j]_q$ with $[0]_q!=1$. With this, the $q$-binomials are
\be
\l[\barr{c} n \\ k\earr\r]_q = \dis\f{\l[n\r]_q!}{\l[n-k\r]_q!\,\l[k\r]_q!}
\label{fl-6a}
\ee
for $n \ge k \ge 0$ and 0 otherwise. Going further, the $q$-normal distribution $f_{qN}(x|q)$ \cite{Ismail,Sza-1}, with $x$ being a standardized variable (then $x$ is zero centered with variance unity), is defined as
\be
f_{qN}(x|q) = \dis\frac{\dis\sqrt{1-q} \dis\prod_{k^\pr=0}^{\infty} \l(1-
q^{k^\pr +1}\r)}{2\pi\,\dis\sqrt{4-(1-q)x^2}}\; \dis\prod_{k^\pr=0}^{\infty}
\l[(1+q^{k^\pr})^2 - (1-q) q^{k^\pr} x^2\r]\;.
\label{fl-7}
\ee
The $f_{qN}(x|q)$ is defined for $x$ in the range defined by $\cs(q)$ where
\be
\cs(q) = \l(-\dis\frac{2}{\dis\sqrt{1-q}}\;,\;+\dis\frac{2}{\dis\sqrt{1-q}}\r)\;.
\label{fl-8}
\ee
with $q$ taking values $0$ to $1$ (in this paper). Note that $f_{qN}(x|q) = 0$ for $x$ outside $\cs(q)$ and the integral of $f_{qN}(x|q)$ is unity, i.e. $\int_{\cs(q)} f_{qN}(x|q)\,dx =1$. For $q=1$, taking the limit properly
will give $f_{qN}(x|1)= (1/\sqrt{2\pi})\,\exp-x^2/2$, the Gaussian with $\cs(q=1)=(-\infty , \infty)$. Also, 
$f_{qN}(x|0)=(1/2\pi) \sqrt{4-x^2}$, the semi-circle with $\cs(q=0)=(-2,2)$. If we put back the centroid $\epsilon$ and the width $\sigma$ in $f_{qN}$, then $\cs(q)$ changes to
$$
\cs(q:\epsilon,\sigma) = \l(\epsilon -\dis\f{2\sigma}{\sqrt{1-q}}\;,\;\epsilon +\dis\f{2\sigma}{\sqrt{1-q}}\r)\;.
$$ 
All odd central moments of $f_{qN}$ are zero and then
the lowest shape parameter is excess or kurtosis $\gamma_2$ that
is simply related to the reduced fourth central moment $\mu_4$,
$\gamma_2=\mu_4-3$. For $f_{qN}$ we have $\mu_4=2+q$. Thus, $\mu_4$ (or $\gamma_2$) determines the value of $q$ \cite{qMK-1}; see Eq. (\ref{fl-qq}) ahead. 

The $q$-Hermite polynomials $He_n(x|q)$, that are orthogonal with $f_{qN}$ as the weight function, are defined by the recursion relation
\be
x\,He_n(x|q) = He_{n+1}(x|q) + \l[n\r]_q\,He_{n-1}(x|q)
\label{fl-9}
\ee
with $He_0(x|q)=1$ and $He_{-1}(x|q)=0$. Note that for $q=1$, the $q$-Hermite
polynomials reduce to normal Hermite polynomials (related to Gaussian) and for
$q=0$ they will reduce to Chebyshev polynomials (related to semi-circle). The polynomials up to order 4 for example are,
\be
\barr{rcl}
He_0(x|q) & = & 1\;,\\
He_1(x|q) & = & x\;,\\
He_2(x|q) & = & x^2-1\;,\\
He_3(x|q) & = & x^3-(2+q)x\;,\\
He_4(x|q) & = & x^4-(3+2q+q^2)x^2+(1+q+q^2)\;.
\earr \label{fl-10}
\ee
Orthogonal property of $He_n(x|q)$'s that plays an important role in the discussion that follows, is
\be
\dis\int^{2/\sqrt{1-q}}_{-2/\sqrt{1-q}} He_n(x|q)\,He_m(x|q)\,f_{qN}(x|q)\,dx = \l[n\r]_q!\,\delta_{mn}\;.
\label{fl-11}
\ee
Using Eq. (\ref{fl-11}), it is easy to derive formulas for the lower order moments of $f_{qN}$. 

With the ensemble averaged eigenvalue density $\overline{\rho(E)}$ for EGOE($k$) or EGUE($k$) being $f_{qN}(E)$, we can seek an expansion of the eigenvalue density $\rho(E)$ of the members of the ensemble in terms of the polynomial excitations of $f_{qN}(E)$ with the polynomials being obviously $q$-Hermite polynomials. This will allow us to study the two-point correlation function and we will turn to this in the following Section. A similar study was made recently \cite{Verb4} for the two-point correlation function in the SYK model.

\section{Eigenvalue density in terms of $q$-Hermite polynomials and the covariances of expansion coefficients determining two-point function}
\label{sec3}

\subsection{Two-point function in terms of $q$-Hermite polynomials}

Eigenvalue density $\rho(E)$ for various members of an embedded random matrix ensemble can be expanded in terms of $q$-Hermite polynomials starting with $q$-normal giving, 
\be
\rho(E)\,dE = f_{qN}(\eh|q) \l[1+ \dis\sum_{\zeta \ge 1}^{\infty} S_\zeta\;\f{He_\zeta(\eh |q)}{\l[\zeta\r]_q!} \r]\,d\eh\;; \;\;\eh=(E-E_c)/\sigma\;.
\label{fl-12}
\ee
Here, $S_\zeta$ are the expansion coefficients and the $S_\zeta$ should not be confused with $S^\rho(x,y)$ used for the two-point function. It is important to recall, as mentioned at the end of Section \ref{sec2} B, the ensemble averaged eigenvalue density $\overline{\rho(E)}$ for EGUE($k$) is $f_{qN}$, i.e
\be
\overline{\rho(E)}\,dE = \sigma^{-1} f_{qN}(\eh)\,d\eh\;.
\label{fl-eerho}
\ee
Therefore in Eq. (\ref{fl-12}), $E_c$ is the centroid and $\sigma$ is the width of $\overline{\rho(E)}$. 
Now, using the expansion given by Eq. (\ref{fl-12}) the distribution function is
\be
F(E) = F_{qN}(E) + d\;\dis\sum_{\zeta \ge 1} \dis\f{S_\zeta}{\l[\zeta\r]_q!}\;\dis\int_{-2/\sqrt{1-q}}^{\eh} f_{qN}(\eh^\pr |q)\,He_{\zeta}(\eh^\pr |q)\,d\eh^\pr
\label{fl-14} 
\ee
and Eqs. (\ref{fl-1}) and(\ref{fl-eerho}) give
\be
\overline{F(E)} = F_{qN}(E) = d\;\dis\int_{-2/\sqrt{1-q}}^{\eh} f_{qN}(\eh^\pr |q)\,d\eh^\pr\;.
\label{fl-15}
\ee
In the limits $q=1$ (i.e. for Gaussians or in the limit $k << m$) and $q=0$ (semi-circle or $k=m$ limit) the integrals in Eqs. (\ref{fl-14}) and (\ref{fl-15}) are easy to obtain. However, for 
a general value of $q$, i.e. for any $k$ value, formulas for these integrals are not known to the best of the author's knowledge. Hence, they need to be evaluated numerically. 
More importantly, the $S_\zeta$ in Eqs. (\ref{fl-12}) and (\ref{fl-14}) are for a given member of the EE($k$) ensemble and it is easy to see that the ensemble average of $S_\zeta$ is zero, i.e. $\overline{S_\zeta}=0$. However, $\overline{S_\zeta S_{\zeta^\prime}} \neq 0$ \cite{Verb4} and these determine the two-point function as discussed ahead. Before turning to this, let us add that in the past, using Eq. (\ref{fl-12}) with additional approximations, some aspects of the variance of level motion in embedded ensembles has been studied by many groups \cite{MF,Br-81,Haq,Leclair,Patel,Chavda}. 

Eq. (\ref{fl-12}) generates an expansion of the two-point function $S^\rho(x,y)$ in terms of $q$-Hermite polynomials (in the reminder of this paper, the symbols $x$ and $y$ are standardized variables, i.e. they denote $\eh$), 
\be
S^{\rho}(x,y) = f_{qN}(x|q)\,f_{qN}(y|q)\,\dis\sum_{\zeta\,,\,\zeta^\pr = 1}^{\infty}
\overline{S_\zeta\,S_{\zeta^\pr}}\;\f{He_\zeta(x |q)}{\l[\zeta\r]_q!}\;\f{He_{\zeta^\pr}(y |q)}{\l[\zeta^\pr \r]_q!}\;.
\label{fl-16}
\ee
Here, it is significant to note that the covariances $\overline{S_\zeta\,S_{\zeta^\pr}}$ of the $S_\zeta$'s are related to the bivariate moments $\widetilde{\Sigma}_{PQ}$ of the two-point function and this is seen as follows. Firstly,
\be
\lan H^p\ran = \overline{\lan H^p\ran} + \dis\sum_{\zeta \ge 1} S_\zeta\,\dis\f{\sigma^p}{\l[\zeta\r]_q!}\;\dis\int_{S(q)} \,x^p\,f_{qN}(x|q)\,He_\zeta(x|q)\,dx \;.
\label{fl-17}
\ee
Note that $\sigma^2=\Sigma_{2,0}=\Sigma_{0,2}$. Now, writing $x^p$ in terms of $q$-Hermite polynomials using 'Proposition 1' in \cite{Sza-2} and then applying Eq. (\ref{fl-11}) will simplify Eq. (\ref{fl-17}) giving,
\be
\barr{l}
\lan H^p\ran = \overline{\lan H^p\ran} + \dis\sum_{\zeta \ge 1} S_\zeta\,\sigma^p\,C_{\f{p-\zeta}{2},p}(q)\;;\\
C_{m,n}(q) = (1-q)^{-m}\;\dis\sum_{j=0}^{m} (-1)^j q^{j(j+1)/2} \l\{\binom{n}{m-j} - \binom{n}{m-j-1}\r\}\;\l[\barr{c} n-2m+j \\ j \earr\r]_q\;.
\earr \label{fl-18}
\ee
This combined with Eq. (\ref{fl-16}) generates formulas for the reduced bivariate moments ${\hat{\Sigma}}_{PQ}$ in terms of the covariances $\overline{S_\zeta\,S_{\zeta^\pr}}$, 
\be
\barr{l}
{\hat{\Sigma}}_{PQ} = \dis\f{{\widetilde{\Sigma}}_{PQ}}{\l[\Sigma_{2,0}\r]^{(P+Q)/2}} = \dis\sum_{\zeta , \zeta^\pr = 1}^{\infty}
\overline{S_\zeta\,S_{\zeta^\pr}}\;C_{\f{P-\zeta}{2},P}(q)
C_{\f{Q-\zeta^\pr }{2},Q}(q)\;.
\earr \label{fl-19}
\ee
Note that $\widetilde{\Sigma}_{Pq}$ is defined by Eq. (\ref{fl-5}) and $\Sigma_{PQ}$ by Eq. (\ref{fl-51a}). 
Let us add that ${\hat{\Sigma}}_{PQ}=0$ for $P+Q$ odd and similarly $\overline{S_\zeta\,S_{\zeta^\pr}}=0$ for $\zeta +\zeta^\pr$ is odd. Also, ${\hat{\Sigma}}_{P0}=0$, ${\hat{\Sigma}}_{PQ} =
{\hat{\Sigma}}_{QP}$, $\overline{S_\zeta}=0$ and  $\overline{S_\zeta\,S_{\zeta^\pr}}=\overline{S_{\zeta^\pr}\,S_{\zeta}}$.

\subsection{Covariances $\overline{S_\zeta\,S_{\zeta^\pr}}$}

Using Eq. (\ref{fl-19}), successively with $P+Q$ increasing from 2, it is easy to see that the covariances $\overline{S_\zeta\,S_{\zeta^\pr}}$ can be written in terms of the moments ${\hat{\Sigma}}_{PQ}$. Formulas for the moments can be derived for low values of $P+Q$ and, as presented in Section \ref{sec4}, at present we can go up to $P+Q=8$ (there are some restrictions for $P+Q=6$ and $8$). With this, $\overline{S_\zeta\,S_{\zeta^\pr}}$ for $\zeta+\zeta^\pr \le 8$ are,
\be
\barr{rcl}
\overline{S_1\,S_1} & = & {\hat{\Sigma}}_{11}\;,\\
\overline{S_3\,S_1} & = & {\hat{\Sigma}}_{31} - C_{13}\,{\hat{\Sigma}}_{11}\;,\\
\overline{S_2\,S_2} & = & {\hat{\Sigma}}_{22}\;,\\
\overline{S_5\,S_1} & = & {\hat{\Sigma}}_{51} - C_{15}\,\overline{S_3\,S_1} - C_{25}\,\overline{S_1\,S_1}\;,\\
\overline{S_4\,S_2} & = & {\hat{\Sigma}}_{42} - C_{14}\,\overline{S_2\,S_2}\;,\\
\overline{S_3\,S_3} & = & {\hat{\Sigma}}_{33} - C^2_{13}\,\overline{S_1\,S_1} - 2 C_{13}\,\overline{S_1\,S_3}\;,\\
\overline{S_7\,S_1} & = & {\hat{\Sigma}}_{71} - C_{17}\,\overline{S_5\,S_1} - C_{27}\,\overline{S_3\,S_1} -
C_{37}\,\overline{S_1\,S_1} \;,\\
\overline{S_6\,S_2} & = & {\hat{\Sigma}}_{62} - C_{16}\,\overline{S_4\,S_2} - C_{26}\,\overline{S_2\,S_2}\;,\\
\overline{S_5\,S_3} & = & {\hat{\Sigma}}_{53} - C_{13}\,\overline{S_5\,S_1} - C_{15}\,\overline{S_3\,S_3}
- \l[C_{25} + C_{15} C_{13}\r]\,\overline{S_1\,S_3}
- C_{25} C_{13}\,\overline{S_1\,S_1}\;,\\
\overline{S_4\,S_4} & = & {\hat{\Sigma}}_{44} - 2C_{14}\,\overline{S_2\,S_4} - C^2_{14}\,\overline{S_2\,S_2}\;.
\earr \label{cov-1}
\ee
In the above, we have dropped $q$ in $C_{mn}(q)$ for brevity. In order to apply Eq. (\ref{cov-1}), Eq. (\ref{fl-18}) for $C_{m,n}(q)$ is simplified for $m=1$, $2$ and $3$ (note that $n \ge 2m+1$). Firstly, $C_{0P}(q)=1$ for any $P$. Similarly, formula for $m=1$ is simple,
\be
C_{1,P}(q) = \dis\sum_{m=2}^P (m-1) q^{P-m}\;.
\label{cov-2}
\ee
Then, for example $C_{1,2}(q) =1$, $C_{1,3}=q+2$, $C_{1,4}=q^2+2q+3$, $C_{1,5}=q^3+2q^2+3q+4$ and so on. 
Besides this, we need the formulas for $C_{25}$, $C_{26}$, $C_{27}$ and $C_{37}$ for applying Eq. (\ref{cov-1}). These are,
\be
\barr{rcl}
C_{2,5}(q) & = & \l[q^3 + 3q^2 + 6q + 5\r]\;,\\
C_{2,6}(q) & = & \l[q^5 + 3q^4 + 7q^3 + 12q^2 + 13q + 9\r]\;,\\
C_{2,7}(q) & = & \l[q^7 + 3q^6 + 7q^5 + 13q^4 + 21q^3 + 24q^2 + 22q + 14\r]\;,\\
C_{3,7}(q) & = & \l[q^6 + 4q^5 + 10q^4 + 20q^3 + 28q^2 + 28q + 
14 \r]\;.
\earr \label{cov-3}
\ee
It is important to note that $\overline{S_i S_j} = \overline{\lan He_i(H)\ran^m\; \lan He_j(H)\ran^m}$ and this can be used
to verify Eq. (\ref{cov-1}).
Now we will derive formulas for the bivariate moments $\Sigma_{PQ}$, with finite $N$ corrections, so that we can obtain lower order covariances of
the $S_\zeta$'s.

\section{Formulas for lower order bivariate moments of two-point
correlation function}
\label{sec4}

In this Section we will derive formulas for the moments $\Sigma_{PQ}$ (there by for $\hat{\Sigma}_{PQ}$) of the two-point correlation function by restricting to EGUE($k$) for a system of $m$ fermions in $N$ single particle states. As established in \cite{Ko-05}, these will follow from the Wigner-Racah algebra of $U(N)$. For EGUE($k$) Hamiltonians all the $m$-fermion states belong to the totally antisymmetric irreducible representation (irrep) $f_m=\{1^m\}$ of $U(N)$ (note that we are using Young tableaux notation for irreps; see Appendix A). Then, the conjugate irrep is $\overline{f_m}=\{1^{N-m}\}$. Given a $k$-body $H$, it will decompose into $U(N)$ tensors $B^\nu(k)$ with the irreps $\nu = \{2^\nu 1^{N-2\nu}\}$; note that $\nu = \overline{\nu}$ (the 'bar' used here for denoting conjugate irrep should not be confused with the 'bar' used for ensemble averages). As $SU(N)$ instead of $U(N)$ is used in the derivations, $\nu =0$ corresponds to $\{1^N\} = \{0\}$ irrep.
With $m$ particle states denoted by $\l.\l|f_m,\alpha\r.\ran$,
we need the $SU(N)$ Clesh-Gordan (CG) coefficients $\lan f_m \alpha_1 \;\;\overline{f_m}\,\overline{\alpha_2} \mid \nu\; \omega_\nu\ran$
where $\alpha$'s and $\omega_\nu$ are additional labels needed for complete specification of various states. In the following we will often use the short hand notation $C^{\nu , \omega_\nu}_{\alpha_1\;\overline{\alpha_2}}$ by dropping the $f_m$ label as always we will deal with $m$-particle states. Some important properties of the CG coefficients $C^{f_{ab} v_{ab}}_{f_a v_a\;f_b v_b} = \lan f_a v_a\;\;f_b v_b \mid f_{ab} v_{ab}\ran$ are \cite{Ko-05,Butler,Butler1},
\be
\barr{l}
C^{0,0}_{\alpha_1\;\overline{\alpha_1}} = \dis\f{1}{\dis\sqrt{d(f_m)}}\;,\;\;\;C^{\nu , \omega_\nu}_{\alpha_2\;\overline{\alpha_1}} = \l\{C^{\nu , \omega_\nu}_{\alpha_1\;\overline{\alpha_2}}\r\}^*\;,\\
\dis\sum_{\alpha_1 , \alpha_2} C^{\nu_1 , \omega_{\nu_1}}_{\alpha_1\;\overline{\alpha_2}}\;
\l\{C^{\nu_2 , \omega_{\nu_2}}_{\alpha_1\;\overline{\alpha_2}}\r\}^* = \delta_{\nu_1 \nu_2}\;\delta_{\omega_{\nu_1} \omega_{\nu_2}}\;,\;\;\;
\dis\sum_{\alpha_1} C^{\nu , \omega_{\nu}}_{\alpha_1\;\overline{\alpha_1}}\;
\l\{C^{0,0}_{\alpha_1\;\overline{\alpha_1}}\r\}^* = \delta_{\nu ,0} \;,\\
C^{f_{ab} v_{ab}}_{f_a v_a\;f_b v_b} = (-1)^{\phi(f_a, f_b, f_{ab})} \,
C^{f_{ab} v_{ab}}_{f_b v_b\;f_a v_a}\;,\;\;\;
C^{f_{ab} v_{ab}}_{f_a v_a\;f_b v_b} = C^{\overline{f_{ab}} \overline{v_{ab}
}}_{\overline{f_a} \overline{v_a}\;\overline{f_b} \overline{v_b}} \;,\\
C^{f_{ab} v_{ab}}_{f_a v_a\;f_b v_b} = (-1)^{\phi(f_a, f_b, f_{ab})}\;
\dis\sqrt{\dis\frac{d(f_{ab})}{d(f_a)}}\;
C^{f_{a} v_{a}}_{f_{ab} v_{ab}\;\overline{f_b} \overline{v_b}} \;.
\earr \label{fl-20}
\ee
Here $d(f)$ is the $U(N)$ dimension of the irrep $\{f\}$ and formula for this is well known \cite{Wy-70}. We have for example $d(f_m) =\binom{N}{m}$. Also $\phi(f_a , f_b , f_{ab}) = \Theta(f_a) + \Theta(f_b) + \Theta(f_{ab})$ and in the present work we do not need the explicit form of the function $\Theta$. Just as the Wigner or CG coefficients, one can define the Racah coefficients for $SU(N)$ \cite{Butler,Butler1}. The Wigner and Racah (or $U-$) coefficients and their various properties will allow one to derive the following important results for the ensemble average of the product any two $m$-particle matrix elements $\lan f_m \alpha_1 \mid H \mid f_m \alpha_2\ran$ of $H$. As proved in \cite{Ko-05} we have,
\be
\barr{rcl}
\overline{H_{\alpha_1 \alpha_2}H_{\alpha_3 \alpha_4}} & = &
\overline{\lan f_m \alpha_1 \mid H \mid f_m \alpha_2\ran\;  
\lan f_m \alpha_3 \mid H \mid f_m \alpha_4\ran} \\
& = & \dis\sum_{\nu = 0,1,\ldots,k; \omega_\nu} \Lambda^\nu(N,m,m-k)\;C^{\nu , \omega_{\nu}}_{\alpha_1\;\overline{\alpha_2}}\;C^{\nu , \omega_{\nu}}_{\alpha_3\;\overline{\alpha_4}}\;,
\earr \label{fl-21}
\ee
and
\be
\overline{H_{\alpha_1 \alpha_2}H_{\alpha_3 \alpha_4}} = \dis\sum_{\mu=0,1,\ldots,m-k; \omega_\mu} \Lambda^\mu(N,m,k) \;C^{\mu , \omega_{\mu}}_{\alpha_1\;\overline{\alpha_4}}\;C^{\mu , \omega_{\mu}}_{\alpha_3\;\overline{\alpha_2}}
\label{fl-22}
\ee
with
\be
\Lambda^{\nu}(N,m,r) = \binom{m-\nu}{r}\,\binom{N-m+r-\nu}{r}\;.
\label{fl-23}
\ee
These equations are important as we use 'binary correlation approximation'. In this approximation, in the ensemble averages involving sums of product of many particle matrix elements of the $H$ operator (similarly any other operator) only terms with pair wise correlated parts will dominate \cite{MF,BRW,Ko-05,SM}.  
Eqs. (\ref{fl-20}), (\ref{fl-21}), (\ref{fl-22}) and (\ref{fl-23}) along with the 'binary correlation approximation' are used to derive formulas for $\Sigma_{PQ}$ and hence for $\hat{\Sigma}_{PQ}$. Now, we will present the results for $\hat{\Sigma}_{PQ}$ with $P+Q = 2$, $4$, $6$ and $8$.

\subsection{Formulas for $\hat{\Sigma}_{PQ}$ with $P+Q=2$}

Formulas for $\Sigma_{2,0} = \Sigma_{0,2}$ and $\Sigma_{1,1}$ are
already presented in \cite{Ko-05} and they are briefly discussed here for completeness. Firstly, the variance $\Sigma_{2,0}$ is simply,
\be
\barr{rcl}
\Sigma_{2,0} & = & \overline{\lan H^2\ran^m} = \dis\f{1}{d(f_m)}
\dis\sum_{\alpha_1, \alpha_2} \overline{H_{\alpha_1 \alpha_2}H_{\alpha_2 \alpha_1}} \\
& = & \dis\f{1}{d(f_m)} \dis\sum_{\mu=0,1,\ldots,m-k; \omega_\mu} \dis\sum_{\alpha_1, \alpha_2} \Lambda^\mu(N,m,k)
\;C^{\mu , \omega_{\mu}}_{\alpha_1\;\overline{\alpha_1}}\;C^{\mu , \omega_{\mu}}_{\alpha_2\;\overline{\alpha_2}} \\
& = & \Lambda^0(N,m,k)\;.
\earr \label{fl-24}
\ee
Here, in the second step we have used Eq. (\ref{fl-22}) and in the last step the fact that $C^{0,0}_{\alpha , \overline{\alpha}}=1/\sqrt{d(f_m)}$ and the sum rules given in Eq. (\ref{fl-20}). Similarly, the $\Sigma_{1,1}$ or the covariance in the eigenvalue centroids is,
\be
\barr{rcl}
\Sigma_{1,1} & = & \overline{\lan H\ran^m \lan H\ran^m} = \dis\f{1}{\l[d(f_m)\r]^2}
\dis\sum_{\alpha_1, \alpha_2} \overline{H_{\alpha_1 \alpha_1}H_{\alpha_2 \alpha_2}} \\
& = & \dis\f{1}{\l[d(f_m)\r]^2} \Lambda^0(N,m,m-k) \dis\sum_{\alpha_1, \alpha_2}
\;C^{0,0}_{\alpha_1\;\overline{\alpha_1}}\;C^{0,0}_{\alpha_2\;\overline{\alpha_2}} \\
& = & \dis\f{1}{d(f_m)}\,\Lambda^0(N,m,m-k)\;.
\earr \label{fl-24a}
\ee
Here, in the second step we have used the result that only a $SU(N)$ scalar ($\nu=0$) part of $H$ contributes to the eigenvalue centroids and also Eq. (\ref{fl-21}). In the last step used is the result $C^{0,0}_{\alpha , \overline{\alpha}}=1/\sqrt{d(f_m)}$ and the sum over $\alpha$'s will give $[d(f_m)]^2$. Combining Eqs. (\ref{fl-24}) and (\ref{fl-24a}) will give the formula for $\hat{\Sigma}_{11}$,
\be
\hat{\Sigma}_{11} = \dis\f{\Lambda^0(N,m,m-k)}{d(f_m)\,\Lambda^0(N,m,k)}\;.
\label{fl-24b}
\ee

\subsection{Formulas for $\hat{\Sigma}_{PQ}$ with $P+Q=4$}

With $P+Q=4$, we have $\Sigma_{4,0} = \Sigma_{0,4}$, $\Sigma_{3,1} = \Sigma_{1,3}$ and $\Sigma_{2,2}$. For $\Sigma_{4,0}$, 
\be 
\Sigma_{4,0} = \overline{\lan H^4\ran^m} = \dis\frac{1}{d(f_m)} \dis\sum_{\alpha_1 , \alpha_2,\alpha_3,\alpha_4} \overline{H_{\alpha_1 \alpha_2} H_{\alpha_2 \alpha_3}H_{\alpha_3 \alpha_4}H_{\alpha_4 \alpha_1}} 
\label{fl-25}
\ee
using binary correlation approximation, there will be two binary correlated terms. Denoting the correlated pairs as $A$, $B$ etc. and applying the cyclic invariance of $m$-particle averages, the two terms are $2 \overline{\lan AA BB \ran^m} = 2 [\overline{\lan AA\ran^m}]^2$ and $\overline{\lan ABAB \ran^m}$. Then, Eq. (\ref{fl-25}) simplifies to
\be
\Sigma_{4,0} = \overline{\lan H^4\ran^m} = 2 \l[\Sigma_{2,0}\r]^2 + \dis\frac{1}{d(f_m)} 
\dis\sum_{\alpha_1 , \alpha_2,\alpha_3,\alpha_4} \overline{H_{\alpha_1 \alpha_2} H_{\alpha_3 \alpha_4}}\;\;\overline{H_{\alpha_2 \alpha_3} H_{\alpha_4 \alpha_1}} 
\label{fl-26}
\ee
Simplifying the last binary correlated term $\overline{\lan ABAB\ran^m}$ using Eqs. (\ref{fl-20}), (\ref{fl-21}) and (\ref{fl-22}) and properties of $SU(N)$ Racah coefficients, we have \cite{Ko-05} (see also \cite{BRW}),
\be
\overline{\lan ABAB\ran^m} = \dis\frac{1}{d(f_m)} \dis\sum_{\nu=0}^{min\{k,m-k\}}
\;\Lambda^\nu(N,m,m-k)\;\Lambda^\nu(N,m,k)\;d(\nu)\;.
\label{fl-27a}
\ee
where $d(\nu) = {\binom{N}{\nu}}^2 - {\binom{N}{\nu-1}}^2$. Then,
\be
\Sigma_{4,0} = \overline{\lan H^4\ran^m} = 2\l[\Lambda^0(N,m,k)\r]^2 + \overline{\lan ABAB\ran^m}\;.
\label{fl-27}
\ee
with the last term given by Eq. (\ref{fl-27a}). Note that Eq. (\ref{fl-27a}) also gives the formula for the $q$ parameter for EGUE($k$) \cite{qMK-1},
\be
q =  \dis\frac{\dis\sum_{\nu=0}^{min\{k,m-k\}}\;\Lambda^\nu(N,m,m-k)\;\Lambda^\nu(N,m,k)\;d(\nu)}{d(f_m)\;\l[\Lambda^0(N,m,k)\r]^2}\;.
\label{fl-qq}
\ee

Turning to $\Sigma_{3,1}$,
\be
\Sigma_{3,1}=\Sigma_{1,3} = \overline{\lan H\ran^m \lan H^3\ran^m}\;,
\label{fl-28a}
\ee
clearly the $H$ matrix element in $\lan H\ran^m$ has to correlate with one of the $H$ matrix elements in $\lan H^3\ran^m$ in the binary correlation approximation. Denoting the correlated terms again as $A$, $B$ etc. we have the three terms $\overline{\lan A\ran^m \lan ABB\ran^m}$, $\overline{\lan A\ran^m \lan BAB\ran^m}$,$\overline{\lan A\ran^m \lan BBA\ran^m}$ and all these three are same due to the cyclic invariance of $m$-particle averages. Then, we have the simple result
\be
\Sigma_{3,1}=\Sigma_{1,3} = 3 \overline{\lan H\ran^m \lan H \ran^m}\;\;\overline{\lan H^2\ran^m} = \dis\f{3}{d(f_m)}\,\Lambda^0(N,m,k)\,\Lambda^0(N,m,m-k)
\label{fl-28}
\ee
and the ensemble averages here follow from Eq. (\ref{fl-24}) and (\ref{fl-24a}). With this $\hat{\Sigma}_{31}$ is,
\be
\hat{\Sigma}_{31} = 3\; \hat{\Sigma}_{11}\;.
\label{fl-28b}
\ee 
Finally, let us consider $\Sigma_{2,2}$,
\be
\Sigma_{2,2} = \overline{\lan H^2\ran^m \lan H^2\ran^m}\;.
\label{fl-29}
\ee
Here again there will be three correlated terms $\overline{\lan AA\ran^m\,\lan BB\ran^m}$, $\overline{\lan AB\ran^m\,\lan AB \ran^m}$ and $\overline{\lan AB\ran^m\,\lan BA\ran^m}$ with the later two equal due to the cyclic invariance of $m$-particle averages. Simplifying these will give easily \cite{Ko-05,BRW},
\be
\barr{rcl}
\Sigma_{2,2} & = & \overline{\lan H^2\ran^m \lan H^2\ran^m} \\
& = & \l[\overline{\lan H^2\ran^m}\r]^2 + 2 \overline{\lan AB\ran^m \lan AB \ran^m}\;;\\
\overline{\lan AB\ran^m \lan AB \ran^m} & = & \dis\f{1}{\l[d(f_m)\r]^2}\;\dis\sum_{\alpha_1, \alpha_2,\alpha_a,\alpha_b}
\overline{H_{\alpha_1 \alpha_2} H_{\alpha_a \alpha_b}}\;\overline{H_{\alpha_2 \alpha_1} H_{\alpha_b \alpha_a}} \\
& = & \dis\f{1}{\l[d(f_m)\r]^2}\;\dis\sum_{\nu=0,1,\ldots,k}\l[\Lambda^\nu(N,m,m-k)\r]^2 d(\nu)\;.
\earr \label{fl-30}
\ee
Note that the formula for $\overline{\lan H^2\ran^m}$ is given by Eq. (\ref{fl-24}). Using this we have for $\hat{\Sigma}_{22}$,
\be
\hat{\Sigma}_{22} = \dis\f{2\;\dis\sum_{\nu=0}^k\;\l[\Lambda^\nu(N,m,m-k)\r]^2 d(\nu)}{\l[d(f_m)\,\Lambda^0(N,m,k) \r]^2}\;.   
\label{fl-30a}
\ee

\subsection{formulas for $\hat{\Sigma}_{PQ}$ with $P+Q=6$}

With $P+Q=6$, we have $\Sigma_{6,0} = \Sigma_{0,6}$, $\Sigma_{5,1} = \Sigma_{1,5}$, $\Sigma_{4,2} = \Sigma_{2,4}$ and $\Sigma_{3,3}$. For $\Sigma_{6,0}$ with,
\be
\Sigma_{6,0}=\Sigma_{0,6} = \overline{\lan H^6\ran^m}\;,
\label{fl-31}
\ee
there will be 4 different binary correlated terms,
\be
\Sigma_{6,0}= 5 \overline{\lan AABBCC\ran^m} + 6 \overline{\lan AABCBC\ran^m} + 3 \overline{\lan ABACBC\ran^m} + \overline{\lan ABCABC\ran^m}
\label{fl-32}
\ee
and the first two terms follow from Eq. (\ref{fl-24}) giving
\be
\barr{l}
5 \overline{\lan AABBCC\ran^m} = 5 \l[\Lambda^0(N,m,k)\r]^3\;,\\
6 \overline{\lan CCABAB\ran^m} = 6 \Lambda^0(N,m,k) \;\overline{\lan ABAB\ran^m}
\earr \label{fl-33}
\ee
with $\overline{\lan ABAB\ran}$ given by Eq. (\ref{fl-27a}).
Now let us consider the third term, 
\be
\barr{l}
\overline{\lan ABACBC\ran^m} = \dis\f{1}{d(f_m)} \dis\sum_{
\alpha_1, \alpha_2,\alpha_3,\alpha_4,\alpha_5,\alpha_6}
\overline{H_{\alpha_1 \alpha_2} H_{\alpha_3 \alpha_4}}\;\;\overline{H_{\alpha_2 \alpha_3} H_{\alpha_5 \alpha_6}}\;\; \overline{H_{\alpha_4 \alpha_5} H_{\alpha_6 \alpha_1}} \;.
\earr \label{fl-33a}
\ee
Applying Eq. (\ref{fl-22}) to the first and third ensemble averages in Eq. (\ref{fl-33a}) and Eq. (\ref{fl-21}) to the second term will give
\be
\barr{l}
\overline{\lan ABACBC\ran^m} = \dis\f{1}{d(f_m)}\;\dis\sum_{\alpha_1, \alpha_2,\alpha_3,\alpha_4,\alpha_6,\alpha_6} \dis\sum_{\mu_1=0,1,\ldots,m-k; \omega_{\mu_1}} \Lambda^{\mu_1}(N,m,k)
\;C^{\mu_1 , \omega_{\mu_1}}_{\alpha_1\;\overline{\alpha_4}}\;C^{\mu_1 , \omega_{\mu_1}}_{\alpha_3\;\overline{\alpha_2}} \\
\times \dis\sum_{\mu_2=0,1,\ldots,m-k; \omega_{\mu_2}} \Lambda^{\mu_2}(N,m,k)
\;C^{\mu_2 , \omega_{\mu_2}}_{\alpha_4\;\overline{\alpha_1}}\;C^{\mu_2 , \omega_{\mu_2}}_{\alpha_6\;\overline{\alpha_5}} \\
\times \dis\sum_{\nu_1=0,1,\ldots,k; \omega_{\nu_1}} \Lambda^{\nu_1}(N,m,m-k)
\;C^{\nu_1 , \omega_{\nu_1}}_{\alpha_2\;\overline{\alpha_3}}\;C^{\nu_1 , \omega_{\nu_1}}_{\alpha_5\;\overline{\alpha_6}} \;.
\earr \label{fl-33b}
\ee
Now, applying the sum rules for the CG coefficients using Eq. (\ref{fl-20}), the final result is obtained,
\be
\overline{\lan ABACBC\ran^m} = \dis\f{1}{d(f_m)}\;\dis\sum_{\nu=0}^{min(k,m-k)} \Lambda^\nu(N,m,m-k)\,\l[\Lambda^\nu(N,m,k)\r]^2\,d(\nu)\;.
\label{fl-33c}
\ee
We are now left with the term $\overline{\lan ABCABC\ran^m}
$ and this can be written as 
\be
\overline{\lan ABCABC\ran^m} = \dis\f{1}{d(f_m)} \dis\sum_{
\alpha_i, \alpha_j,\alpha_k,\alpha_\ell,\alpha_P,\alpha_Q}
\overline{H_{\alpha_i \alpha_j} H_{\alpha_k \alpha_\ell}}\;\;
\overline{H_{\alpha_j \alpha_P} H_{\alpha_\ell \alpha_Q}}\;\;
\overline{H_{\alpha_P \alpha_k} H_{\alpha_Q \alpha_i}}\;.
\label{fl-33d}
\ee
It is easy to see that Eq. (\ref{fl-33d}) is same as the $S_3$ term in \cite{KM-strn}; see Eq. (32) in this paper. Then, its simplification involves $SU(N)$ Racah (or $U-$) coefficients.
The final result follows from Eq. (36) of \cite{KM-strn} with $t=k$ giving,
\be
\barr{l}
\overline{\lan ABCABC\ran^m} = \dis\f{1}{\l[d(f_m)\r]^2} \dis\sum_{\mu_1,\mu_2=0}^k \dis\sum_{\nu=0}^{
min(2k,m-k)} d(\mu_1)\, d(\mu_2)\; \l|U(f_m \mu_1 f_m \mu_2\,;\,f_m \nu)\r|^2 \\
\times \;\Lambda^{\mu_1}(N,m,m-k)\,\Lambda^{\mu_2}(N,m,m-k)\,\Lambda^{\nu}(N,m,k)\;.
\earr \label{fl-33e}
\ee
In Eq. (\ref{fl-33e}), for simplicity we are not showing the multiplicities that appear in the $U$-coefficient. See \cite{Butler,Butler1} for $SU(N)$ Racah coefficients and for some of
their properties. Eq. (48) of \cite{KM-strn} gives the formula in the asymptotic limit for the $U$-coefficient appearing above and this is used in Appendix C. Eqs. (\ref{fl-32}), (\ref{fl-33}), (\ref{fl-33c}) and (\ref{fl-33e}) together will give the formula for $\Sigma_{6,0} = \Sigma_{0,6}$,
\be
\barr{rcl}
\Sigma_{6,0} & = & 5\;\l[\Lambda^0(N,m,k)\r]^3 + \dis\f{6}{d(f_m)}\;\Lambda^0(N,m,k) \;\dis\sum_{\nu=0}^{min(k,m-k)}
\;\Lambda^\nu(N,m,m-k)\;\Lambda^\nu(N,m,k)\;d(\nu) \\
&+& \dis\f{3}{d(f_m)}\;\dis\sum_{\nu=0}^{min(k,m-k)} \Lambda^\nu(N,m,m-k)\,\l[\Lambda^\nu(N,m,k)\r]^2\,d(\nu) \\
&+& \dis\f{1}{\l[d(f_m)\r]^2} \dis\sum_{\mu_1,\mu_2=0}^k \,\dis\sum_{\nu=0}^{
min(2k,m-k)} d(\mu_1)\, d(\mu_2)\; \l|U(f_m \mu_1 f_m \mu_2\,;\,f_m \nu)\r|^2 \\
&\times& \;\Lambda^{\mu_1}(N,m,m-k)\,\Lambda^{\mu_2}(N,m,m-k)\,\Lambda^{\nu}(N,m,k)\;.
\earr \label{fl-33f}
\ee
Though $\hat{\Sigma}_{6,0}=0$, we need the formula for $\Sigma_{6,0}$ when we consider $\hat{\Sigma}_{PQ}$ with $P+Q \ge 8$; see Subsection D.

Formula for $\Sigma_{5,1}$ is simple and this follows from the same arguments that gave Eq. (\ref{fl-28}). Then,
\be
\Sigma_{5,1} = \Sigma_{1,5} = \overline{\lan H^5\ran^m \lan H \ran^m} = 5\,\overline{\lan H \ran^m \lan H \ran^m}\;\overline{\lan H^4 \ran^m}
\label{fl-34}
\ee
with the first factor given by Eq. (\ref{fl-24a}) and the second factor by Eq. (\ref{fl-27}). Then,
\be
\hat{\Sigma}_{5,1} = 10\; \hat{\Sigma}_{1,1} + \dis\frac{5 \,\hat{\Sigma}_{1,1}\;
\dis\sum_{\nu=0}^{min\{k,m-k\}}
\;\Lambda^\nu(N,m,m-k)\;\Lambda^\nu(N,m,k)\;d(\nu)}{d(f_m)\,\l[\Lambda^0(N,m,k)\r]^2} \;.
\label{fl-34a}
\ee 
Coming to $\Sigma_{4,2}$, it is easy to see that there are three different binary correlation terms giving,
\be
\barr{l}
\Sigma_{4,2} = \Sigma_{2,4} = \overline{\lan H^4 \ran^m \,\lan H^2 \ran^m} \\
= \overline{\lan H^2 \ran^m}\;\overline{\lan H^4 \ran^m} + 8 \overline{\lan ABCC\ran^m \lan AB \ran^m} + 4 \overline{\lan ABCB\ran^m \lan AC\ran^m} \\
= \overline{\lan H^2 \ran^m}\;\overline{\lan H^4 \ran^m} + 8  
\overline{\lan H^2 \ran^m}\;\overline{\lan AB\ran^m \lan AB \ran^m} + 4 \overline{\lan ABCB\ran^m \lan AC\ran^m} \;.
\earr \label{fl-35}
\ee 
Here the first two terms follow from Eqs. (\ref{fl-24}), (\ref{fl-27}) and (\ref{fl-30}) and the third term is simplified as follows. Firstly as in Eq. (\ref{fl-33d}),
\be
\barr{l}
\overline{\lan ABCB\ran^m \lan AC\ran^m}  = \dis\f{1}{\l[d(f_m)\r]^2}\;\dis\sum_{\alpha_1, \alpha_2,\alpha_3,\alpha_4,\alpha_a,\alpha_b}
\overline{H_{\alpha_1 \alpha_2} H_{\alpha_a \alpha_b}}\;\;\overline{H_{\alpha_2 \alpha_3} H_{\alpha_4 \alpha_1}}\;\; \overline{H_{\alpha_3 \alpha_4} H_{\alpha_b \alpha_a}} \\
= \dis\f{1}{\l[d(f_m)\r]^2}\;\dis\sum_{\alpha_1, \alpha_2,\alpha_3,\alpha_4,\alpha_a,\alpha_b}\;\; \dis\sum_{\mu_1=0,1,\ldots,k; \omega_{\mu_1}} \Lambda^{\mu_1}(N,m,m-k)
\;C^{\mu_1 , \omega_{\mu_1}}_{\alpha_1\;\overline{\alpha_2}}\;C^{\mu_1 , \omega_{\mu_1}}_{\alpha_a\;\overline{\alpha_b}} \\
\times \dis\sum_{\mu_2=0,1,\ldots,k; \omega_{\mu_2}} \Lambda^{\mu_2}(N,m,m-k)
\;C^{\mu_2 , \omega_{\mu_2}}_{\alpha_3\;\overline{\alpha_4}}\;C^{\mu_2 , \omega_{\mu_2}}_{\alpha_b\;\overline{\alpha_a}} \\
 \times \dis\sum_{\nu_1=0,1,\ldots,m-k; \omega_{\nu_1}} \Lambda^{\nu_1}(N,m,k)
\;C^{\nu_1 , \omega_{\nu_1}}_{\alpha_2\;\overline{\alpha_1}}\;C^{\nu_1 , \omega_{\nu_1}}_{\alpha_4\;\overline{\alpha_3}} \;.
\earr \label{fl-36}
\ee
Now the sum rules for the CG coefficients as given by Eq. (\ref{fl-20}),  allow us to carry out the sum over all the $\alpha$'s giving $\mu_1 = \mu_2 = \nu_1$ and similarly $\omega_{\mu_1} = \omega_{\mu_2} = \omega_{\nu_1}$. With these,
Eq. (\ref{fl-36}) simplifies to
\be
\overline{\lan ABCB\ran^m \lan AC\ran^m}  = \dis\f{1}{\l[d(f_m)\r]^2}\;\dis\sum_{\nu=0}^{min(k,m-k)} \Lambda^\nu(N,m,k)\,\l[\Lambda^\nu(N,m,m-k)\r]^2\,d(\nu)\;.
\label{fl-37}
\ee
Combining Eqs. (\ref{fl-37}) and (\ref{fl-35}) will give the formula for $\Sigma_{4,2}=\Sigma_{2,4}$. Now, formula for $\hat{\Sigma}_{4,2}$ is
\be
\hat{\Sigma}_{4,2} = 4\; \hat{\Sigma}_{2,2} + \dis\f{4\;\dis\sum_{\nu=0}^{min(k,m-k)} \Lambda^\nu(N,m,k)\,\l[\Lambda^\nu(N,m,m-k)\r]^2\,d(\nu)}{\l[d(f_m)\r]^2\,\l[\Lambda^0(N,m,k)\r]^3}\;.
\label{fl-37a}
\ee

Finally, for $\Sigma_{3,3}$ defined by
\be
\Sigma_{3,3} = \overline{\lan H^3\ran^m \lan H^3\ran^m}
\label{fl-38}
\ee
there will be three binary correlated terms,
\be
\Sigma_{3,3} = 9 \overline{\lan ABB \ran^m \lan ACC\ran^m} + 3 \overline{\lan ABC\ran^m \lan ACB\ran^m} + 3 \overline{\lan ABC\ran^m \lan ABC\ran^m}\;.
\label{fl-39}
\ee
Note that by definition, for EGUE($k$), $\overline{\lan H^3\ran^m}=0$. The first term in Eq. (\ref{fl-39}) is simple,
\be
\overline{\lan ABB\ran^m \lan ACC\ran^m} = \l[\overline{\lan H^2\ran^m}\r]^2\;\overline{\lan H\ran^m \lan H\ran^m}\;.
\label{fl-40}
\ee
The second term in Eq. (\ref{fl-39}) has a structure quite similar to the one in Eq. (\ref{fl-36}),
\be
\overline{\lan ABC\ran^m \lan ACB\ran^m} = \dis\f{1}{\l[d(f_m)\r]^2}\;\dis\sum_{\alpha_1, \alpha_2,\alpha_3,\alpha_a,\alpha_b,\alpha_c}
\overline{H_{\alpha_1 \alpha_2} H_{\alpha_a \alpha_b}}\;\;\overline{H_{\alpha_2 \alpha_3} H_{\alpha_c \alpha_a}}\;\; \overline{H_{\alpha_3 \alpha_1} H_{\alpha_b \alpha_c}}\;.
\label{fl-41}
\ee
Simplifying just as in Eq. (\ref{fl-36}) and (\ref{fl-37}), will
give the final formula,
\be
\overline{\lan ABC\ran^m \lan ACB\ran^m} = \dis\f{1}{\l[d(f_m)\r]^2}\;\dis\sum_{\mu=0,1,\ldots,m-k} \l[\Lambda^\mu(N,m,k)\r]^3\;d(\mu)\;.
\label{fl-42}
\ee
The third term $\overline{\lan ABC\ran^m \lan ABC\ran^m}$ has structure quite similar to $\overline{\lan ABCABC\ran^m}$. Following the same steps that led to Eq. (\ref{fl-33e}) will give the formula involving $SU(N)$ $U$-coefficients. Firstly,
\be
\overline{\lan ABC\ran^m \lan ABC\ran^m} = \dis\f{1}{\l[d(f_m)\r]^2}\;\dis\sum_{\alpha_1, \alpha_2,\alpha_3,\alpha_a,\alpha_b,\alpha_c}
\overline{H_{\alpha_1 \alpha_2} H_{\alpha_a \alpha_b}}\;\;\overline{H_{\alpha_2 \alpha_3} H_{\alpha_b \alpha_c}}\;\; \overline{H_{\alpha_3 \alpha_1} H_{\alpha_c \alpha_a}}\;.
\label{fl-42a}
\ee
Now, simplifying this using the same procedure as in Eqs. (32)-(36) of \cite{KM-strn} will generate the following formula,
\be
\barr{l}
\overline{\lan ABC\ran^m\,\lan ABC\ran^m} = \dis\f{1}{\l[d(f_m)\r]^3} \dis\sum_{\nu,\nu_1,\nu_2 = 0}^k \;d(\nu_1)\, d(\nu_2)\;\l|U(f_m \nu_1 f_m \nu_2\,;\,f_m \nu)\r|^2 \\
\times \Lambda^{\nu_1}(N,m,m-k)\,\Lambda^{\nu_2}(N,m,m-k)\,\Lambda^{\nu}(N,m,m-k) \;.
\earr \label{fl-43}
\ee
As in Eq. (\ref{fl-33e}), again in Eq. (\ref{fl-43}), for simplicity, we are not showing the multiplicities that appear in the $U$-coefficient. Eqs. (\ref{fl-39}), (\ref{fl-40}), (\ref{fl-42}) and (\ref{fl-43}) will give the formula for $\Sigma_{3,3}$. With these, the formula for $\hat{\Sigma}_{3,3}$ is
\be
\barr{l}
\hat{\Sigma}_{3,3} = 9\,\hat{\Sigma}_{1,1} + \dis\f{3\;\dis\sum_{\mu=0}^{m-k} \l[\Lambda^\mu(N,m,k)\r]^3\;d(\mu)}{
\l[d(f_m)\r]^2 \l[\Lambda^0(N,m,k)\r]^3} \\
+ \dis\f{3}{\l[d(f_m)\;\Lambda^0(N,m,k)\r]^3} \dis\sum_{\nu,\nu_1,\nu_2 = 0}^k \;d(\nu_1)\, d(\nu_2)\;\l|U(f_m \nu_1 f_m \nu_2\,;\,f_m \nu)\r|^2 \\
\times \;\Lambda^{\nu_1}(N,m,m-k)\,\Lambda^{\nu_2}(N,m,m-k)\,\Lambda^{\nu}(N,m,m-k) \;.
\earr \label{fl-43a}
\ee
Thus, we have simple finite-$N$ formulas for all $\hat{\Sigma}_{P,Q}$ with $P+Q=6$ except for the $U$-coefficient in Eq. (\ref{fl-43a}).

\subsection{formulas for $\hat{\Sigma}_{PQ}$ with $P+Q=8$}

With $P+Q=8$, we need to derive formulas for $\hat{\Sigma}_{7,1} = \hat{\Sigma}_{1,7}$, $\hat{\Sigma}_{6,2} = \hat{\Sigma}_{2,6}$, $\hat{\Sigma}_{5,3} = \hat{\Sigma}_{3,5}$ and $\hat{\Sigma}_{4,4}$. Firstly, $\hat{\Sigma}_{7,1}$ is simple,
\be
\hat{\Sigma}_{7,1} = \hat{\Sigma}_{1,7} = \dis\f{\overline{
\lan H^7\ran^m \;\lan H \ran^m}}{\l[\Sigma_{2,0}\r]^4} = 7\;\hat{\Sigma}_{11}\;\dis\f{\Sigma_{6,0}}{\l[\Lambda^0(N,m,k)\r]^3}
\label{fl-44}
\ee
and the formula for $\Sigma_{6,0}$ is given by Eq. (\ref{fl-33f}). 

Formula for $\hat{\Sigma}_{6,2}$ is more complicated and $\Sigma_{6,2}$ contains
four different terms,
\be
\barr{l}
\Sigma_{6,2} = \Sigma_{2,6} = \overline{\lan H^6 \ran^m \,\lan H^2 \ran^m} = \overline{\lan H^2 \ran^m}\;\;\overline{\lan H^6 \ran^m} + 12\, \overline{\lan ABH^4\ran^m \lan AB \ran^m} \\
+ 12\, \overline{\lan ABCDEF\ran^m \lan AC\ran^m} + 6\,\overline{\lan ABCDEF\ran^m \lan AD\ran^m} \;.
\earr \label{fl-45}
\ee 
The second term here is simple giving
\be
\overline{\lan ABH^4\ran^m \lan AB\ran^m} = \overline{\lan AB\ran^m \lan AB\ran^m}\;\;\overline{\lan H^4\ran^m}
\label{fl-46}
\ee
and the formulas for the two terms on the R.H.S. are given by Eqs. (\ref{fl-30}), (\ref{fl-27}) and (\ref{fl-27a}). The third term in Eq. (\ref{fl-45}) is,
\be
\barr{l}
\overline{\lan ABCDEF\ran^m \lan AC\ran^m} = \dis\f{1}{\l[d(f_m)\r]^2} \;\dis\sum_{\alpha_1, \alpha_2,\alpha_3,\alpha_4,\alpha_5,\alpha_6,\alpha_a,\alpha_b}\; X_1\;\l[X_2+X_3+X_4\r]\;;\\ 
X_1 = \overline{H_{\alpha_1 \alpha_2} H_{\alpha_a \alpha_b}}\;\;\overline{H_{\alpha_3 \alpha_4} H_{\alpha_b \alpha_a}} \;,\;\;\;
X_2 = \overline{H_{\alpha_2 \alpha_3} H_{\alpha_4 \alpha_5}}\;\;\;\overline{H_{\alpha_5 \alpha_6} H_{\alpha_6 \alpha_1}} \;,\\
X_3 = \overline{H_{\alpha_2 \alpha_3} H_{\alpha_5 \alpha_6}}\;\;
\overline{H_{\alpha_4 \alpha_5} H_{\alpha_6 \alpha_1}} \;,\;\;\;
X_4 = \overline{H_{\alpha_2 \alpha_3} H_{\alpha_6 \alpha_1}}\;\;
\overline{H_{\alpha_4 \alpha_5} H_{\alpha_5 \alpha_6}} \;.
\earr \label{fl-47}
\ee
Firstly, the '$X_1$' term is simplified using Eq. (\ref{fl-21}). Similarly, the second part of $X_2$,i.e. $\overline{H_{\alpha_5 \alpha_6} H_{\alpha_6 \alpha_1}}$ gives $\Lambda^0(N,m,k)\,\delta_{\alpha_1 \alpha_5}$. Then, $X_1 X_2$ with sum over all
the $\alpha$'s, after applying Eq. (\ref{fl-21}), is given by
\be
\barr{l}
\dis\sum_{\alpha's} X_1 X_2 = \dis\sum_{\nu=0}^k \l[\Lambda^\nu(N,m,m-k)\r]^2 \Lambda^0(N,m,k)\;\dis\sum_{\mu=0,1,\ldots,m-k}
\Lambda^\mu(N,m,k) \\
\times \; \dis\sum_{\alpha_1,\alpha_2,\alpha_3, \alpha_4, \omega_\nu, \omega_\mu} C^{\nu , \omega_\nu}_{\alpha_1 \overline{\alpha_2}}\; C^{\nu, \omega_\nu}_{\alpha_3 \overline{\alpha_4}}\;C^{\mu , \omega_\mu}_{\alpha_2 \overline{\alpha_1}}\;
C^{\mu , \omega_\mu}_{\alpha_4 \overline{\alpha_3}} \\
= \Lambda^0(N,m,k)\;\dis\sum_{\nu=0}^{min(k,m-k)} \l[\Lambda^\nu(N,m,m-k)\r]^2 \Lambda^\nu(N,m,k)\;d(\nu)\;.
\earr \label{fl-48}
\ee
In the last step here we have used the sum rules for the CG coefficients as given in Eq. (\ref{fl-20}). Going further it is easy to see that the $X_1 X_4$ with sum over all the $\alpha$'s is same as $X_1 X_2$ with sum over all the $\alpha$'s. Then, we are left with $X_1 X_3$. Applying Eqs. (\ref{fl-21}) and (\ref{fl-22}) will give,
\be
\barr{l}
\dis\sum_{\alpha's} X_1 X_3 = \dis\sum_{\nu=0}^k \l[\Lambda^\nu(N,m,m-k)\r]^2 \;\;\dis\sum_{\mu=0}^k \Lambda^\mu(N,m,m-k)\;\dis\sum_{\mu\pr = 0,1,\ldots,m-k}
\Lambda^{\mu^\pr}(N,m,k) \\
\times\; \dis\sum_{\alpha_1,\alpha_2,\alpha_3, \alpha_4, \alpha_5,\alpha_6, \omega_\nu, \omega_\mu, \omega_{\mu^\pr}}
C^{\nu , \omega_\nu}_{\alpha_1 \overline{\alpha_2}}\; C^{\nu, \omega_\nu}_{\alpha_3 \overline{\alpha_4}}\;C^{\mu , \omega_\mu}_{\alpha_2 \overline{\alpha_3}}\;
C^{\mu , \omega_\mu}_{\alpha_5 \overline{\alpha_6}}\;
C^{\mu^\pr , \omega_{\mu^\pr}}_{\alpha_4 \overline{\alpha_1}}\;
C^{\mu^\pr , \omega_{\mu^\pr}}_{\alpha_6 \overline{\alpha_5}}\;.
\earr \label{fl-49}
\ee
This is simplified using Eq. (\ref{fl-20}) and the transformation of product of two CG coefficients as given by Eq. (21) of \cite{Ko-05}. Applying these will give the formula,
\be
\barr{l}
\dis\sum_{\alpha's} X_1 X_3 = \dis\sum_{\nu=0}^k \dis\sum_{\mu=0}^{min(k,m-k)} \l[\Lambda^\nu(N,m,m-k)\r]^2 \Lambda^\mu(N,m,m-k)
\Lambda^\mu(N,m,k) \\
\times\;\dis\sqrt{d(\nu) d(\mu)}\;U(f_m \overline{f_m} f_m f_m\,;\,\nu \mu)\;.
\earr \label{fl-50}
\ee
Combining Eqs. (\ref{fl-48}) and (\ref{fl-50}) will give the formula for $\overline{\lan ABCDEF\ran^m \lan AC\ran^m}$.
Turning to the fourth term in $\Sigma_{6,2}$, first we have
\be
\barr{l}
\overline{\lan ABCDEF\ran^m \lan AD\ran^m} = \dis\f{1}{\l[d(f_m)\r]^2} \;\dis\sum_{\alpha_1, \alpha_2,\alpha_3,\alpha_4,\alpha_5,\alpha_6,\alpha_a,\alpha_b}\; Y_1\;\l[Y_2+Y_3+
Y_4\r]\;;\\ 
Y_1 = \overline{H_{\alpha_1 \alpha_2} H_{\alpha_a \alpha_b}}\;\;\overline{H_{\alpha_4 \alpha_5} H_{\alpha_b \alpha_a}} \;,\;\;\;
Y_2 = \overline{H_{\alpha_2 \alpha_3} H_{\alpha_3 \alpha_4}}\;\;\;\overline{H_{\alpha_5 \alpha_6} H_{\alpha_6 \alpha_1}} \;,\\
Y_3 = \overline{H_{\alpha_2 \alpha_3} H_{\alpha_5 \alpha_6}}\;\;
\overline{H_{\alpha_3 \alpha_4} H_{\alpha_6 \alpha_1}} \;,\;\;\;
Y_4=\overline{H_{\alpha_2 \alpha_3} H_{\alpha_6 \alpha_1}}\;\;
\overline{H_{\alpha_3 \alpha_4} H_{\alpha_5 \alpha_6}} \;.
\earr \label{fl-51}
\ee
The $Y_1Y_2$ term is simplified easily using Eqs. (\ref{fl-20})-(\ref{fl-22}) and similarly the $Y_1Y_4$ term giving,
\be
\barr{l}
\dis\sum_{\alpha's} Y_1 Y_2 = \dis\sum_{\nu=0}^k \l[\Lambda^\nu(N,m,m-k)\r]^2 \l[\Lambda^0(N,m,k)\r]^2\,d(\nu)\;,\\
\dis\sum_{\alpha's} Y_1 Y_4 = \dis\sum_{\nu=0}^{min(k,m-k)} \l[\Lambda^\nu(N,m,m-k)\;\Lambda^\nu(N,m,k)\r]^2\,d(\nu)\;.
\earr \label{fl-52}
\ee
Simplification of the $Y_1Y_3$ term needs not only Eqs. (\ref{fl-20})-(\ref{fl-22}) but also Eq. (21) of \cite{Ko-05} 
for $Y_1$ and and Eq. (35) of \cite{KM-strn} for $Y_3$. With these we have,
\be
\barr{l}
\dis\sum_{\alpha's} Y_1 Y_3 = \dis\sum_{\mu=0}^{2k} \dis\sum_{\mu_1, \mu_2, \nu=0}^k \l[\Lambda^\nu(N,m,m-k)\r]^2\;\Lambda^{\mu_1}(N,m,m-k)\;\Lambda^{\mu_2}(N,m,m-k) \\
\times\;\dis\f{d(\mu_1) d(\mu_2)\sqrt{d(\nu)}}{d(f_m) \sqrt{d(\mu)}}\;U(f_m \overline{f_m} f_m f_m ; \nu \mu)
\l|U(f_m \mu_1 f_m \mu_2 ; f_m \mu)\r|^2\;.
\earr \label{fl-53}
\ee
Combining Eqs. (\ref{fl-52}) and (\ref{fl-53}) will give the formula for
$\overline{\lan ABCDEF\ran^m \lan AD\ran^m}$. 
With all these, formula for $\hat{\Sigma}_{6,2}$ is
\be
\barr{l}
\hat{\Sigma}_{6,2} = \hat{\Sigma}_{2,6} = \l[\Lambda^0(N,m,k)\r]^{-4}\;
\l\{A1+A2+A3\r\}\;;\\
A1 = \dis\f{12}{\l[d(f_m)\r]^2}\;\dis\sum_{\nu=0}^k\;\l\{\l[\Lambda^\nu(N,m,m-k)\r]^2 d(\nu)\r\}\;\l\{2\l[\Lambda^0(N,m,k)\r]^2 \r.\\ 
\l. + \dis\frac{1}{d(f_m)} \dis\sum_{\nu=0}^{min(k,m-k)}
\;\Lambda^\nu(N,m,m-k)\;\Lambda^\nu(N,m,k)\;d(\nu)\r\}\;,\\
A2 = \dis\f{24}{\l[d(f_m)\r]^2}\;\Lambda^0(N,m,k)\;\dis\sum_{\nu=0}^{min(k,m-k)} \l[\Lambda^\nu(N,m,m-k)\r]^2 \Lambda^\nu(N,m,k)\;d(\nu) \\
+ \dis\f{12}{\l[d(f_m)\r]^2}\;\l\{\;\dis\sum_{\nu=0}^k \dis\sum_{\mu=0}^{min(k,m-k)} \l[\Lambda^\nu(N,m,m-k)\r]^2 \Lambda^\mu(N,m,m-k) \Lambda^\mu(N,m,k) \r. \\
\l. \times\;\dis\sqrt{d(\nu) d(\mu)}\;U(f_m \overline{f_m} f_m f_m\,;\,\nu \mu) \r\}\;,\\
A3 = \dis\f{6}{\l[d(f_m)\r]^2}\;\l[\Lambda^0(N,m,k)\r]^2\;\dis\sum_{\nu=0}^k \l[\Lambda^\nu(N,m,m-k)\r]^2 \;d(\nu) \\
+ \dis\f{6}{\l[d(f_m)\r]^2}\;\dis\sum_{\nu=0}^{min(k,m-k)} \l[\Lambda^\nu(N,m,m-k)\r]^2\;\l[\Lambda^\nu(N,m,k)\r]^2 \;d(\nu) \\
+ \dis\f{6}{\l[d(f_m)\r]^2}\;\l\{\;\dis\sum_{\mu=0}^{2k} \dis\sum_{\mu_1, \mu_2, \nu=0}^k \l[\Lambda^\nu(N,m,m-k)\r]^2\;\Lambda^{\mu_1}(N,m,m-k)\;\Lambda^{\mu_2}(N,m,m-k) \r. \\
\l. \times\;\dis\f{d(\mu_1) d(\mu_2)\sqrt{d(\nu)}}{d(f_m) \sqrt{d(\mu)}}\;U(f_m \overline{f_m} f_m f_m ; \nu \mu)
\l|U(f_m \mu_1 f_m \mu_2 ; f_m \mu)\r|^2\r\} \;.
\earr \label{fl-58}
\ee
Now, we will consider $\hat{\Sigma}_{5,3}$ and $\hat{\Sigma}_{4,4}$. 

It is easy to see that $\hat{\Sigma}_{5,3}$ and $\hat{\Sigma}_{4,4}$ will involve much larger number of terms than $\Sigma_{6,2}$ and they will also involve several $SU(N)$ $U$-coefficients. Details of various terms in $\hat{\Sigma}_{5,3}$ and $\hat{\Sigma}_{4,4}$ are given in Appendix B. Following this, the formula for $\hat{\Sigma}_{5,3}$ is 
\be
\hat{\Sigma}_{5,3} = 15(2+q)\,\hat{\Sigma}_{1,1} + 5(1+q) \l[\hat{\Sigma}_{3,3} -9 \hat{\Sigma}_{1,1}\r] + X_{53}\;.
\label{fl-s53-a1}
\ee    
Here, used are Eqs. (\ref{s53-a}), (\ref{s53-2}), (\ref{fl-42}), (\ref{fl-43}),
(\ref{fl-43a}) and (\ref{s53-4}) with $X_{53}$ defined by Eq. (\ref{s53-4}). Similarly, formula for $\hat{\Sigma}_{4,4}$ is
\be
\barr{l}
\hat{\Sigma}_{4,4} = 16 \hat{\Sigma}_{2,2} + 8 \l[\hat{\Sigma}_{4,2} -4 \hat{\Sigma}_{2,2}\r]  \\
+ \dis\f{8}{\l[d(f_m)\r]^2} \dis\sum_{\nu=0}^{min(k,m-k)}
\dis\f{\l[\Lambda^{\nu}(N,m,m-k)\,\Lambda^{\nu}(N,m,k)\r]^2\,d(\nu)}{\l[\Lambda^0(N,m,k)\r]^4} \\
+ \dis\f{4}{\l[d(f_m)\r]^2} \dis\sum_{\nu=0}^{m-k}
\dis\f{\l[\Lambda^{\nu}(N,m,k)\r]^4\,d(\nu)}{\l[\Lambda^0(N,m,k)\r]^4} + X_{44}\;.
\earr \label{fl-s44-a1}
\ee
Here, used are Eqs. (\ref{s44-2}), (\ref{fl-24}), (\ref{fl-30}), (\ref{fl-37}, (\ref{s44-4}) and (\ref{s44-8}) with $X_{44}$ defined by Eqs. (\ref{s44-5}) and (\ref{s44-6}). Note that
$X_{53}$ in Eq. (\ref{fl-s53-a1}) and $X_{44}$ in Eq. (\ref{fl-s44-a1}) involve
$SU(N)$ $U$-coefficients for which formulas are not available. However, both
$X_{53}$ and $X_{44}$ can be neglected in the asymptotic limit (see Appendix C).

Formulas derived in this Section along with Eq. (\ref{cov-1}) will allow us to calculate $\overline{S_iS_j}$ numerically for $i+j \le 8$ as well as allow to examine their asymptotic structure. We will turn to these in the following Section.
  
\begin{table}
\caption{Covariances $\overline{S_iS_j}$ for EGUE($k$). Results are shown for $(N,m)=(12,6)$ with $k=2$ and $3$ and for $(N,m)=(30,10)$ for $k=2$, $3$ and $4$. Numbers in the brackets are from asymptotic limit formulas given by Eq. (\ref{eq.fsisj}). Given also are the $q$ values obtained using Eq. (\ref{fl-qq}).}
\label{table1}
{\footnotesize{
\begin{tabular}{c|c|c|c}  
\hline
& $N=12,\;\;m=6\;\;\;$ & & \\
$\overline{S_iS_j}\;\;\;$ & $k=2\;\;\;$ & $k=3$ & \\
\hline
& $q=0.287$ & $q=0.006$ & \\
$\overline{S_1S_1}$ & $8.117 \times 10^{-3}$ ($3.444 \times 10^{-3}$) & $1.082 \times 10^{-3}$ ($4.132 \times 10^{-4}$) & \\
$\overline{S_3S_1}$ & $5.787 \times 10^{-3}$ ($3.444 \times 10^{-3}$) & $1.021 \times 10^{-3}$ ($4.132 \times 10^{-4}$) \\
$\overline{S_2S_2}$ & $1.160 \times 10^{-3}$ ($4.591 \times 10^{-4}$) & $1.227 \times 10^{-4}$ ($4.132 \times 10^{-5}$) & \\
$\overline{S_5S_1}$ & $6.835 \times 10^{-3}$ ($4.845 \times 10^{-3}$) & $1.075 \times 10^{-3}$ ($3.968 \times 10^{-4}$) \\
$\overline{S_4S_2}$ & $1.533 \times 10^{-3}$ ($5.251 \times 10^{-4}$) & $1.497 \times 10^{-4}$ ($4.064 \times 10^{-5}$) & \\
$\overline{S_3S_3}$ & $3.062 \times 10^{-2}$ ($1.303 \times 10^{-2}$) & $5.176 \times 10^{-3}$ ($1.974 \times 10^{-3}$) & \\
$\overline{S_7S_1}$ & $1.249 \times 10^{-2}$ ($1.063 \times 10^{-2}$) & $1.282 \times 10^{-3}$ ($3.993 \times 10^{-4}$) & \\
$\overline{S_6S_2}$ & ($3.676 \times 10^{-4}$) & ($4.276 \times 10^{-5}$) \\
$\overline{S_5S_3}$ & ($1.539 \times 10^{-2}$) &  ($2.08 \times 10^{-3}$) \\
$\overline{S_4S_4}$ & ($3.907 \times 10^{-4}$) & ($4.126 \times 10^{-5}$) \\
\hline
& $N=30,\;\;m=10\;\;\;$ & & \\
$\overline{S_iS_j}\;\;\;$ & $k=2\;\;\;$ & $k=3\;\;\;$ & $k=4$ \\
\hline
& $q=0.524$ & $q=0.208$ & $q=0.005$ \\
$\overline{S_1S_1}$ & $4.478 \times 10^{-4}$ ($2.378 \times 10^{-4}$) & $1.669 \times 10^{-5}$ ($7.28 \times 10^{-6}$) & $7.211 \times 10^{-7}$ ($2.796 \times 10^{-7}$) \\
$\overline{S_3S_1}$ & $2.13 \times 10^{-4}$ ($2.378 \times 10^{-4}$) & $1.322 \times 10^{-5}$ ($7.28 \times 10^{-6}$) &
$6.884 \times 10^{-7}$ ($2.796 \times 10^{-7}$) \\
$\overline{S_2S_2}$ & $1.718 \times 10^{-5}$ ($1.057 \times 10^{-5}$) & $2.577 \times 10^{-7}$ ($1.213 \times 10^{-7}$) & 
$6.719 \times 10^{-9}$ ($2.663 \times 10^{-9}$) \\
$\overline{S_5S_1}$ & $2.354 \times 10^{-4}$ ($2.413 \times 10^{-4}$) & $1.528 \times 10^{-5}$ ($9.725 \times 10^{-6}$) &
$7.182 \times 10^{-7}$ ($3.151 \times 10^{-7}$) \\
$\overline{S_4S_2}$ & $1.559 \times 10^{-5}$ ($9.731 \times 10^{-6}$) & $3.025 \times 10^{-7}$ ($1.365 \times 10^{-7}$) & 
$7.532 \times 10^{-9}$ ($2.797 \times 10^{-9}$) \\
$\overline{S_3S_3}$ & $1.177 \times 10^{-3}$ ($6.252 \times 10^{-4}$) & $6.887 \times 10^{-5}$ ($3.004 \times 10^{-5}$) & 
$3.473 \times 10^{-6}$ ($1.345\times 10^{-6}$) \\
$\overline{S_7S_1}$ & $5.852 \times 10^{-4}$ ($6.316 \times 10^{-4}$) & $2.016 \times 10^{-5}$ ($1.826 \times 10^{-5}$) & 
$7.659 \times 10^{-7}$ ($3.885 \times 10^{-7}$) \\
$\overline{S_6S_2}$ & ($1.703 \times 10^{-6}$) & ($1.141 \times 10^{-7}$) &
($2.746 \times 10^{-9}$) \\
$\overline{S_5S_3}$ & ($6.91 \times 10^{-4}$) &  ($3.472 \times 10^{-5}$) &
($1.405 \times 10^{-6}$)
\\
$\overline{S_4S_4}$ & ($5.566 \times 10^{-6}$) & ($1.111 \times 10^{-7}$) &
($2.652 \times 10^{-9}$)
\\
\hline
\end{tabular}
}}
\end{table}

\section{Asymptotic limit results for the covariances and expansion for the number variance}
\label{sec5}

In the previous Section we have derived formulas for $\hat{\Sigma}_{P,Q}$ with
$P+Q=2-8$. In particular, the formula for $(P,Q)=(1,1)$ is given by Eq.
(\ref{fl-24b}), for $(3,1)$ by Eq. (\ref{fl-28b}), for $(2,2)$ by Eq.
(\ref{fl-30a}), for $(5,1)$ by Eq. (\ref{fl-34a}), for $(4,2)$ by Eq.
(\ref{fl-37a}), for $(3,3)$ by Eq. (\ref{fl-43a}), for $(7,1)$ by Eqs.
(\ref{fl-44}) and (\ref{fl-33f}), for $(6,2)$ by Eq. (\ref{fl-58}), for $(5,3)$
by Eq. (\ref{fl-s53-a1}) and finally for $(4,4)$ by Eq. (\ref{fl-s44-a1}). 
Also, note that formula for the $q$ parameter is given by Eq. (\ref{fl-qq}) and
$\Lambda^\nu(N,m,r)$ is given by Eq. (\ref{fl-23}). In addition, the dimensions
$d(f_m)=\binom{N}{m}$ and $d(\nu) = {\binom{N}{\nu}}^2 - {\binom{N}{\nu-1}}^2$.
Using all these equations along with Eq. (\ref{cov-1}), the covariances
$\overline{S_iS_j}$ for $(i,j)=(1,1)$, $(3,1)$, $(2,2)$, $(5,1)$, $(4,2)$,
$(3,3)$ and $(7,1)$ are calculated and the results are shown in Table
\ref{table1}. For $(3,3)$, the last term in Eq. (\ref{fl-43a}) is not included
as the $U$-coefficients needed here are not available. Finite $N$ results for
$(6,2)$ are not shown as formulas for the two $U$-coefficients appearing in Eq.
(\ref{fl-37a}) are not available. Similarly, the finite $N$ results for $(5,3)$ and $(4,4)$ are not shown in the Table. It is seen from the Table that in general, the covariances are small and they are of the same order of magnitude as in the SYK model (for Majorana fermions) reported earlier in \cite{Verb4}. 

\subsection{Asymptotic limit formulas for $\overline{S_iS_j}$}

For further insight into the structure of $\overline{S_iS_j}$, the formulas in
Section \ref{sec4} are used to derive asymptotic limit formulas for $\sh_{PQ}$
and these are given in Appendix C. Now, using the formulas in Eq. (\ref{asy4}) and Eqs.
(\ref{cov-1})-(\ref{cov-3}), the following asymptotic limit formulas are obtained for  $\overline{S_iS_j}$ with $i+j \le 8$,
\be 
\barr{rcl}
\overline{S_1 S_1} & = & \dis\f{\binom{m}{k}}{{\binom{N}{k}}^2}\;,\\
\overline{S_3 S_1} & = & (1-q)\, \overline{S_1 S_1} \;,\\
\overline{S_2 S_2} & = & \dis\f{1}{{\binom{N}{k}}^2}\;,\\
\overline{S_5 S_1} & = & (1-q^2)^2\, \overline{S_1 S_1} \;,\\
\overline{S_4 S_2} & = & (1-q^2)\, \overline{S_2 S_2} \;,\\
\overline{S_3 S_3} & = & 3(1-q)\, \overline{S_1 S_1} + \dis\f{3}{\binom{m}{k} \binom{N}{k}^2} + O\l(\f{1}{\binom{N}{k}^4}\r) \;,\\
\overline{S_7 S_1} & = & (1-q)(1-q^2)^2\l[1+2q+3q^3+2q^3+q^4 \r]\, \overline{S_1 S_1} \;,\\
\overline{S_6 S_2} & = & \l(q^6+q^5-q^4+4q^3-7q^2+q+1\r)\, \overline{S_2 S_2} + O\l(\f{1}{\binom{N}{k}^4}\r)\;,\\
\overline{S_5 S_3} & = & (1-q)^2\l[q^3 + 7q^2 +11q +5)\r]\,\overline{S_1 S_1} + \dis\f{3(1-q)(q^2+3q+1)}{\binom{m}{k} \binom{N}{k}^2}  + O\l(\f{1}{\binom{N}{k}^4}\r)\;,\\
\overline{S_4 S_4} & = & \l(1-q^2\r)^2\,\overline{S_2 S_2}  +
\dis\f{4}{\binom{m}{k}^2\,\binom{N}{k}^2} + O\l(\f{1}{\binom{N}{k}^4}\r) \;;\\
q & = & \dis\f{\binom{m-k}{k}}{\binom{m}{k}}\;.
\earr \label{eq.fsisj}
\ee
All these formulas agree with the GUE ($m=k$ giving $q=0$) results given in \cite{Br-81,FMP}. Also, they agree with results for EGUE($k$) in the $k <<m$ limit as given in \cite{MF,Br-81} and this corresponds to $q=1$ in Eq. (\ref{eq.fsisj}). Thus, the results in Eq. (\ref{eq.fsisj}) apply to all $k$ values ranging from from $k << m$ to $k=m$. Also, the results in Section \ref{sec4} give finite $N$ corrections to the formulas in Eq. (\ref{eq.fsisj}) for all $k$ values. 

Going further, numerical results given by Eq. (\ref{eq.fsisj}) are shown in brackets in Table \ref{table1}. These results are not too far from the finite $N$ results. The correlations, as seen from the asymptotic limit formulas in Eq. (\ref{eq.fsisj}) are of the order of $1/\l[\binom{N}{k}\r]^2$.
With the $1/\l[\binom{N}{k}\r]^2$ scaling, correlations $\overline{S_i S_j}$
shown in Table \ref{table1} are no longer small. More strikingly, for $q
\rightarrow 1$ (i.e. $k/m \rightarrow 0$) the $\overline{S_i S_j}=0$ for $i \neq j$
and $\overline{S^2_i}$ = $\binom{m}{k}^{2-i} \,\binom{N}{k}^{-2}$. Similarly, for $q=0$ (i.e. $k=m$) the structure of $\overline{S_i S_j}$ is simple and $\overline{S_i S_j} \neq 0$ both for $i=j$ and $i \neq j$. However, for intermediate $k$ values (between $k << m$ and $k=m$), $\overline{S_i S_j}$ are a combination of $q$, $\overline{S_1S_1}$, $\overline{S_2S_2}$ and $\binom{m}{k}^{r} \,\binom{N}{k}^{-2}$ with $r \leq 1$. 

Numerical  evaluation of $\Sigma_{6,0}$, $\Sigma_{3,3}$ and $\Sigma_{6,2}$ (also for $\Sigma_{5,3}$ and $\Sigma_{4,4}$ - see Appendix B) requires formulas for $SU(N)$ $U$-coefficients
of the type $U(f_m \overline{f_m} f_m f_m ; \nu \mu)$ and $U(f_m \nu_1 f_m \nu_2 ; f_m \nu)$. Neither the formulas nor a viable method for their determination is available in literature. The situation here is similar to the $U$-coefficients needed even for the fourth moment for EGUE's with
spin and spin-isospin $SU(4)$ symmetries \cite{Ko-07,KM-10} as encountered before. Thus, much of the progress in analytical approach to EGUE($k$)'s will depend on extending our knowledge on
$SU(N)$ $U$-coefficients. One approach is to develop further the so called pattern calculus introduced by Louck and Biedenharn many years back for $SU(N)$ Wigner-Racah algebra \cite{BidLou-1,BidLou-2,BidLou-3}. Another is to derive asymptotic expansions for the $SU(N)$ Racah coefficients as attempted in the past by French \cite{Sun-asmp}. 

\subsection{Expansion for number variance $\Sigma^2(\on)$}

Before concluding the paper, as an example it is instructive to
consider the expansion for the number variance $\Sigma^2(\on)$
in terms of $\overline{S_iS_j}$ and this follows from the expansion for the two-point function. The definition given by Eq. (\ref{fl-3}) together with Eqs. (\ref{fl-2}) and (\ref{fl-14}) will
give the expansion,
\be
\barr{l}
\Sigma^2(\on) = d^2 \dis\sum_{\zeta , \zeta^\prime =1}^{\infty}\;
\overline{S_\zeta \,S_{\zeta^\prime}}\; \l[R_\zeta(x|q) - R_\zeta(y|q)\r]\l[R_{\zeta^\prime}(x|q)-R_{\zeta^\prime}(y|q)\r]\;;\\
\\
R_{\zeta}(x|q) = \dis\int_{-\f{2}{\sqrt{1-q}}}^x f_{qN}(z|q) \dis\f{He_\zeta(
z|q)}{[\zeta]_q!}\,dz\;.
\earr \label{exnum}
\ee
Note that we have used the property $\overline{S_\zeta}=0$. In Eq. (\ref{exnum}), with $\Sigma^2(\on)$ defined over $x_0 \pm (\on \overline{D})/2$, $x=x_0 - (\on \overline{D})/2$ and $y=x_0 + (\on \overline{D})/2$. Note that $\overline{D}$ is average mean spacing (in $\sigma$ units) and $x_0$ is the eigenvalue around which $\Sigma^2(\on)$ is evaluated. It is expected that $\Sigma^2(\on)$ to be independent on $x_0$ except perhaps near the spectrum ends. With formulas for $\overline{S_iS_j}$ for $i+j \leq 8$ available as given by the equations in Section \ref{sec4} along with Eq. (\ref{cov-1}), the series given by Eq. (\ref{exnum}) can be evaluated up to $\zeta+\zeta^\prime \leq 8$ terms. Alternatively one can use the asymptotic limit formulas given by Eq. (\ref{eq.fsisj}). Note that at present the function
$R_\zeta(x|q)$ need to be evaluated numerically as no analytical formula for the integral defining $R_\zeta(x|q)$ is available except for $q=1$ and $0$.   

Finally, direct derivation of asymptotic limit formulas for many other $\sh_{PQ}$ for higher $P+Q$ values (i.e. $P+Q > 8$) may prove to be useful in future as they will provide systematics for $\hat{\Sigma}_{PQ}$ and hence
for $\overline{S_iS_j}$. With this, it may be possible to carry out the sum in Eq. (\ref{fl-16}) (or the sum in Eq. (\ref{exnum})) and obtain the two-point function (or the number variance) for EGUE($k$) just as it was carried out using the moment method for GOE and GUE in the past \cite{Br-81,Ko-book,FMP}. This work is left for future.

\section{Conclusions and future outlook}
\label{sec6}

Two-point correlation function in eigenvalues of embedded random matrix ensembles with $k$-body interactions is not yet available though these ensembles are applied to many different quantum systems in the last 50 years (see \cite{Ko-book,KC-rev1,KC-rev2,MaSe,Ben1,Ben2,Ben3,Ben4,Iz-1,Iz-2,KM-annals} and references there in
for the past and for more recent applications of EE). With the recent recognition
that the one-point function for these ensembles follows $q$-normal form, it is possible to seek an expansion of the eigenvalue density of the members of the ensemble in terms of
$q$-Hermite polynomials. Covariances $\overline{S_\zeta S_{\zeta^\pr}}$ of the expansion coefficients $S_\zeta$ with $\zeta \ge 1$ here determine the two-point function. As the covariances are linear combination of the bivariate moments $\Sigma_{PQ}$ of the two-point function (see Section \ref{sec3}), in this paper, in Section \ref{sec4},
derived are formulas for the bivariate moments $\Sigma_{PQ}$ with $P+Q \le 8$ for the embedded Gaussian unitary ensembles with $k$-body interactions [EGUE($k$)] as appropriate for systems with $m$ fermions in $N$ single particle states. The Wigner-Racah algebra for $SU(N)$ plays a central role in deriving the formulas with finite $N$ corrections \cite{Ko-05,KM-strn}. However, the $\Sigma_{PQ}$ with $P+Q=6$ and 8 need extension of the available knowledge in calculating $SU(N)$ $U$-coefficients; see Section \ref{sec4}. Using the finite $N$ formulas, in Section \ref{sec5} derived are asymptotic limit ($N \rightarrow \infty$, $m \rightarrow \infty$, $m/N \rightarrow 0$ with $k$ finite) formulas for $\overline{S_\zeta S_{\zeta^\pr}}$ with $\zeta+\zeta^\prime \leq 8$. 

In future, expecting the availability of new methods for evaluating general $SU(N)$ $U$-coefficients, it may be possible to get systematics of $\Sigma_{PQ}$ and $\overline{S_\zeta S_{\zeta^\pr}}$ and with these it may be possible to derive the two-point correlation function for EGUE($k$) ensemble [perhaps also for EGOE($k$) and EGSE($k$)]. Once the two-point function is available, this may also open the possibility of studying ergodicity and stationarity properties of EGUE($k$); See \cite{MF,Br-81,Selig} for some past attempts in this direction.  

\acknowledgments

Thanks are due to N.D. Chavda and Manan Vyas for some useful correspondence. 

\renewcommand{\theequation}{A-\arabic{equation}}
\setcounter{equation}{0}   
\section*{APPENDIX A}

Reduction of the Kronecker product of the irreps $\nu_1$ and $\nu_2$ giving irreps $\nu_3$ is symbolically denoted by 
\be
\nu_1 \times \nu_2 = \dis\sum_{\nu_3} \Gamma_{\nu_1 \nu_2 \nu_3} \;\nu_3
\label{app1}
\ee
where $\times$ denotes Kronecker product and $\Gamma$ gives the multiplicity, i.e. number of times $\nu_3$ appears in the Kronecker product. If $\Gamma_{\nu_1 \nu_2 \nu_3}=0$ implies that the irrep $\nu_3$ will not appear in the Kronecker product. 
In our applications, the irreps $\nu$ correspond to the Young tableaux $\{2^{\nu} 1^{N-2\nu}\}$ of $U(N)$. Then, Eq. (\ref{app1}) changes to
\be
\l\{2^{\nu_1} 1^{N-2\nu_1}\r\} \times \l\{2^{\nu_2} 1^{N-2\nu_2}\r\} = \dis\sum_{\nu_3} \Gamma_{\nu_1 \nu_2 \nu_3} \l\{2^{\nu_3} 1^{N-2\nu_3}\r\} \;.
\label{app2}
\ee
Though the methods to obtain the reduction given by Eq. (\ref{app2}) are well known \cite{Wy-70,Little}, a simpler approach is to first evaluate the Kronecker product of the transpose of the irreps and then take the transpose of the final irreps. By taking transpose, the two column irreps $\{2^{\nu} 1^{N-2\nu}\}$ change to two rowed irreps $\{N-\nu , \nu\}$ giving
\be
\l\{N-\nu_1 , \nu_1\r\} \times \l\{N-\nu_2 , \nu_2\r\} = \dis\sum_{\nu_3} \Gamma_{\nu_1 \nu_2 \nu_3}\, \l\{N-\nu_3 , \nu_3\r\}\;.
\label{app3}
\ee
The Kronecker product here is easy to evaluate using the identity
\be
\l\{N-\nu_1 , \nu_1\r\} \times \l\{N-\nu_2 , \nu_2\r\} =      
\l\{N-\nu_1 , \nu_1\r\} \times \l[ \l\{N-\nu_2\r\} \times \l\{\nu_2\r\} -  \l\{N-\nu_2 +1\r\} \times \l\{\nu_2 -1\r\}\r]\;.
\label{app4}
\ee
Now, the product $\{n_1,n_2\} \times \{n_3\}$ is simply sum of the irreps $\{n_1+n_a,n_2+n_b,n_c\}$ with $n_a \ge 0$, $n_b \le n_1-n_2$, $n_c \le n_2$ and $n_a+n_b+n_c = n_3$. Similarly, for the product $\{n_1,n_2,n_3\} \times \{n_4\}$; see \cite{Wy-70,Little} and Eq. (B.9) in \cite{Ko-su3}. Applying this to Eq. (\ref{app4}) gives 
in general 2, 3 and 4 rowed irreps, however we need only two rowed irreps. Regularization of the 3 and 4 rowed irreps is done using the rules: (i) four rowed irreps $\{n_1,n_2,n_3,n_4\}=0$ if $n_1 \ne N$ and $n_2 \ne N$. As $n_1+n_2+n_3+n_4=2N$, the allowed irrep is just $\{N,N,0,0\}$; (ii) three rowed irreps $\{n_1,n_2,n_3\} = \{n_2,n_3\}$ if $n_1=N$ and $0$ otherwise. Also, note that $\nu=0$ corresponds to $\{1^N\}$ for $U(N)$ and $\{0\}$ for $SU(N)$. Using all these, we find for $N >> \nu$ and $N$ large the following results,
\be
\barr{rcl}
\nu \times 1 & = & (\nu \pm 1)^1 , (\nu)^2 \;,\\
\nu \times 2 & = & (\nu \pm 2)^1 , (\nu \pm 1)^2 , (\nu)^3 \;,\\
\nu \times 3 & = & (\nu \pm 3)^1 , (\nu \pm 2)^2 , (\nu \pm 1)^3, (\nu)^4\;,\\
\nu \times 4 & = & (\nu \pm 4)^1 , (\nu \pm 3)^2 , (\nu \pm 2)^3, (\nu \pm 1)^4, (\nu)^5\;.
\earr \label{app5}
\ee
In the above, $r$ in $(\mu)^r$ denotes multiplicity of the irrep
$\mu$. Continuing the above for $\nu \times 5$, $\nu \times 6$ etc., it is easy to see that $\nu \times \nu$ always gives the irrep $\nu$ but with multiplicity.

\renewcommand{\theequation}{B-\arabic{equation}}
\setcounter{equation}{0}   
\section*{APPENDIX B}

Let us consider $\Sigma_{5,3}$,
\be
\Sigma_{5,3} = \Sigma_{3,5} = \overline{\lan H^5\ran^m\,\lan H^3\ran^m}\;.
\label{s53-1}
\ee
Firstly, $\overline{\lan H^3\ran^m}=0$ and $\overline{\lan H^5\ran^m}=0$ for EGUE($k$). Therefore,
\be
\hat{\Sigma}_{5,3} = \hat{\Sigma}_{3,5} = \l[\overline{\lan H^2\ran^m}\r]^{-4}\,\Sigma_{5,3} 
\label{s53-a}
\ee
In the binary correlation approximation, for $\Sigma_{5,3}$ there are two possibilities: (i) one $H$ in $\lan H^3\ran^m$ correlates with one of the $H$'s in $\lan H^5\ran^m$; (ii) the three $H$'s in $\lan H^3\ran^m$ correlate pairwise with three of the $H$'s in $\lan H^5\ran^m$. These will give 5 binary correlated terms,
\be
\barr{rcl} 
\Sigma_{5,3} & = & 15 \overline{\lan A H^4\ran^m\;\lan A H^2\ran^m} + 15 \overline{\lan ABC DD\ran^m\;\lan ABC\ran^m} \\
& + & 15 \overline{\lan ABC DD\ran^m\;\lan ACB\ran^m} +
15 \overline{\lan ABDCD\ran^m\;\lan ABC\ran^m} + 15 \overline{\lan ABDCD\ran^m\;\lan ACB\ran^m} \\
& = & 15 \overline{\lan H\ran^m\;\lan H \ran^m}\;\;
\overline{\lan H^4\ran^m}\;\;\overline{\lan H^2\ran^m} + 15
\overline{\lan H^2\ran^m}\;\l[\overline{\lan ABC\ran^m \lan ABC\ran^m} + \overline{\lan ABC\ran^m \lan ACB\ran^m}\r] \\
& + & 15 \overline{\lan ABDCD\ran^m\;\lan ABC\ran^m} + 15 \overline{\lan ABDCD\ran^m\;\lan ACB\ran^m} \;.
\earr \label{s53-2}
\ee
Except for the last two terms, formulas for rest of the terms
in Eq. (\ref{s53-2}) are already given in Section \ref{sec4}; see Eqs. (\ref{fl-42}) and (\ref{fl-43}). The last two terms are,
\be
\barr{l}
\overline{\lan ABDCD\ran^m \lan ABC\ran^m} = \dis\f{1}{\l[d(f_m)\r]^2} \;\dis\sum_{\alpha_1, \alpha_2,\alpha_3,\alpha_4,\alpha_5,\alpha_a,\alpha_b,\alpha_c}\; 
\overline{H_{\alpha_1 \alpha_2} H_{\alpha_a \alpha_b}}\;\;\overline{H_{\alpha_2 \alpha_3} H_{\alpha_b \alpha_c}} \\
\times \;\;\overline{H_{\alpha_3 \alpha_4} H_{\alpha_5 \alpha_1}}\;\;\overline{H_{\alpha_4 \alpha_5} H_{\alpha_c \alpha_a}} \;,\\
\\
\overline{\lan ABDCD\ran^m \lan ACB\ran^m} = \dis\f{1}{\l[d(f_m)\r]^2} \;\dis\sum_{\alpha_1, \alpha_2,\alpha_3,\alpha_4,\alpha_5,\alpha_a,\alpha_b,\alpha_c}\; 
\overline{H_{\alpha_1 \alpha_2} H_{\alpha_a \alpha_b}}\;\;\overline{H_{\alpha_2 \alpha_3} H_{\alpha_c \alpha_a}} \\
\times \;\;\overline{H_{\alpha_3 \alpha_4} H_{\alpha_5 \alpha_1}}\;\;\overline{H_{\alpha_4 \alpha_5} H_{\alpha_b \alpha_c}} 
\;.
\earr \label{s53-3}
\ee
Further simplification of these follow from the $SU(N)$ algebra given in
\cite{Ko-05,KM-strn,Butler,Butler1}. Clearly, these will involve several $SU(N)$
$U$-coefficients. However, assuming $N \rightarrow \infty$ and $(m,k)$ finite,
it is possible to use the approximation $\lan ABDCD\ran^m \sim
[\binom{m}{k}]^{-1} \binom{m-k}{k}\;\lan ABCDD\ran^m$; see \cite{MF,Ko-book}.
Also, $q \sim [\binom{m}{k}]^{-1} \binom{m-k}{k}$. With these, the last two
terms in Eq. (\ref{s53-2}) can be written as
\be
\barr{l}
15\overline{\lan ABDCD\ran^m \lan ABC\ran^m} + 15\overline{\lan ABDCD\ran^m \lan ACB\ran^m} \\
= 15\,q\;\overline{\lan H^2\ran^m}\;
\l[\overline{\lan ABC\ran^m \lan ABC\ran^m} + \overline{\lan ABC\ran^m \lan ACB\ran^m}\r] + X_{53}
\earr \label{s53-4}
\ee
Here, $X_{53}$ is the correction to the approximation given by the first term
and this is expected to be of the order of $\l[\binom{N}{k}\r]^{-4}$. With this, $\Sigma_{5,3}$ is 
\be
\barr{rcl}
\Sigma_{5,3} & = & 15\, \overline{\lan H\ran^m\;\lan H \ran^m}\;\; \overline{\lan H^4\ran^m}\;\;\overline{\lan H^2\ran^m} \\
& + & 15(1+q)\;
\overline{\lan H^2\ran^m}\;\l[\overline{\lan ABC\ran^m \lan ABC\ran^m} + \overline{\lan ABC\ran^m \lan 
ACB\ran^m}\r] + X_{53}\;.
\earr \label{s53-4a}
\ee 

Turning to $\Sigma_{4,4}$,
\be
\Sigma_{4,4} = \overline{\lan H^4\ran^m\,\lan H^4\ran^m}\;,
\label{s44-1}
\ee 
in the binary correlation approximation, there are three possibilities: (i) the two $\lan H^4\ran^m$'s are independent; (ii) two of $H$'s in one $\lan H^4\ran^m$ correlate with two $H$'s in the other $\lan H^4\ran^m$; (iii) the four $H$'s in $\lan H^4\ran^m$ correlate pairwise with the four $H$'s in the other $\lan H^4\ran^m$. These will give 1, 3 and 6 binary correlated terms respectively,
\be
\barr{l}
\Sigma_{4,4} = \overline{\lan H^4\ran^m}\;\;\overline{\lan H^4\ran^m} + 32 \overline{\lan ABCC\ran^m\;\lan ABDD\ran^m} + 32
\overline{\lan ABCC\ran^m\;\lan ADBD\ran^m} \\
+ 8 \overline{\lan ACBC\ran^m\;\lan ADBD\ran^m}  
+ 4 \overline{\lan ABCD\ran^m\;\lan ABCD\ran^m} 
+ 4 \overline{\lan ABCD\ran^m\;\lan ABDC\ran^m} \\
+ 4 \overline{\lan ABCD\ran^m\;\lan ACBD\ran^m} 
+ 4 \overline{\lan ABCD\ran^m\;\lan ACDB\ran^m} \\
+ 4 \overline{\lan ABCD\ran^m\;\lan ADBC\ran^m} 
+ 4 \overline{\lan ABCD\ran^m\;\lan ADCB\ran^m} \;.
\earr \label{s44-2}
\ee
Formula for the first term is given by Eqs. (\ref{fl-27}) and (\ref{fl-27a}). Further, the second term reduces to $\l(\overline{\lan H^2\ran^m}\r)^2\; \overline{\lan AB\ran^m \lan AB\ran^m}$ and formula for this follows from Eqs. (\ref{fl-24}) and (\ref{fl-30}). Similarly, the third term reduces to $\overline{\lan H^2\ran^m}\;\overline{\lan AB\ran^m \lan ADBD\ran^m}$ and the formula for this follows from 
Eq. (\ref{fl-37}). The fourth term is explicitly,
\be
\barr{l}
\overline{\lan ACBC\ran^m \lan ADBD\ran^m} = \dis\f{1}{\l[d(f_m)\r]^2} \;\dis\sum_{\alpha_1, \alpha_2,\alpha_3,\alpha_4,\alpha_a,\alpha_b,\alpha_c,\alpha_d}\; 
\overline{H_{\alpha_1 \alpha_2} H_{\alpha_a \alpha_b}}\;\;\overline{H_{\alpha_2 \alpha_3} H_{\alpha_4 \alpha_1}} \\
\times \;\;\overline{H_{\alpha_3 \alpha_4} H_{\alpha_c \alpha_d}}\;\;\overline{H_{\alpha_b \alpha_c} H_{\alpha_d \alpha_a}} \\
= \dis\f{1}{\l[d(f_m)\r]^2}\;\dis\sum_{\alpha_1,
\alpha_2,\alpha_3,\alpha_4,\alpha_a,\alpha_b,\alpha_c,\alpha_d}\,\dis\sum_{\nu_1,\nu_3=0}^k
\,\dis\sum_{\nu_2,\nu_4=0}^{m-k} \dis\sum_{\omega_{\nu_1},\omega_{\nu_2},\omega_{\nu_3},\omega_{\nu_4}}     \\
\Lambda^{\nu_1}(N,m,m-k)\, C^{\nu_1 , \omega_{\nu_1}}_{\alpha_1
\overline{\alpha_2}} C^{\nu_1, \omega_{\nu_1}}_{\alpha_a
\overline{\alpha_b}}\;\Lambda^{\nu_2}(N,m,k)\, C^{\nu_2 , \omega_{\nu_2}}_{\alpha_2 \overline{\alpha_1}}\;
C^{\nu_2 , \omega_{\nu_2}}_{\alpha_4 \overline{\alpha_3}} \\
\Lambda^{\nu_3}(N,m,m-k)\, C^{\nu_3 , \omega_{\nu_3}}_{\alpha_3
\overline{\alpha_4}} C^{\nu_3, \omega_{\nu_3}}_{\alpha_c
\overline{\alpha_d}}\;\Lambda^{\nu_4}(N,m,k)\, C^{\nu_4 , \omega_{\nu_4}}_{\alpha_b \overline{\alpha_a}}\;
C^{\nu_4 , \omega_{\nu_4}}_{\alpha_d \overline{\alpha_c}} \;.
\earr \label{s44-3}
\ee
Here we have applied Eqs. (\ref{fl-21}) and (\ref{fl-22}). Now, simplifying the CG coefficients will give the formula,
\be
\overline{\lan ACBC\ran^m \lan ADBD\ran^m} = \dis\f{1}{\l[d(f_m)\r]^2}
\dis\sum_{\nu=0}^{min(k,m-k)} \l[\Lambda^{\nu}(N,m,m-k)\,\Lambda^{\nu}(N,m,k)\r]^2\,d(\nu)\;.
\label{s44-4}
\ee  
For the last 6 terms in Eq. (\ref{s44-2}) we can write formulas similar to the one in Eq. (\ref{s44-3}). Firstly, the first five terms are,
\be
\barr{l}
\overline{\lan ABCD\ran^m \lan ABCD\ran^m} = X_1 = \dis\f{1}{\l[d(f_m)\r]^2} \;\dis\sum_{\alpha_1, \alpha_2,\alpha_3,\alpha_4,\alpha_a,\alpha_b,\alpha_c,\alpha_d}\; 
\overline{H_{\alpha_1 \alpha_2} H_{\alpha_a \alpha_b}}\;\;\overline{H_{\alpha_2 \alpha_3} H_{\alpha_b \alpha_c}} \\
\times \;\;\overline{H_{\alpha_3 \alpha_4} H_{\alpha_c \alpha_d}}\;\;\overline{H_{\alpha_4 \alpha_1} H_{\alpha_d \alpha_a}} \;,\\
\overline{\lan ABCD\ran^m \lan ABDC\ran^m} = X_2 = \dis\f{1}{\l[d(f_m)\r]^2} \;\dis\sum_{\alpha_1, \alpha_2,\alpha_3,\alpha_4,\alpha_a,\alpha_b,\alpha_c,\alpha_d}\; 
\overline{H_{\alpha_1 \alpha_2} H_{\alpha_a \alpha_b}}\;\;\overline{H_{\alpha_2 \alpha_3} H_{\alpha_b \alpha_c}} \\
\times \;\;\overline{H_{\alpha_3 \alpha_4} H_{\alpha_d \alpha_a}}\;\;\overline{H_{\alpha_4 \alpha_1} H_{\alpha_c \alpha_d}} \;,\\
\overline{\lan ABCD\ran^m \lan ACBD\ran^m} = X_3 = \dis\f{1}{\l[d(f_m)\r]^2} \;\dis\sum_{\alpha_1, \alpha_2,\alpha_3,\alpha_4,\alpha_a,\alpha_b,\alpha_c,\alpha_d}\; 
\overline{H_{\alpha_1 \alpha_2} H_{\alpha_a \alpha_b}}\;\;\overline{H_{\alpha_2 \alpha_3} H_{\alpha_c \alpha_d}} \\
\times \;\;\overline{H_{\alpha_3 \alpha_4} H_{\alpha_b \alpha_c}}\;\;\overline{H_{\alpha_4 \alpha_1} H_{\alpha_d \alpha_a}} \;,\\
\overline{\lan ABCD\ran^m \lan ACDB\ran^m} = X_4 = \dis\f{1}{\l[d(f_m)\r]^2} \;\dis\sum_{\alpha_1, \alpha_2,\alpha_3,\alpha_4,\alpha_a,\alpha_b,\alpha_c,\alpha_d}\; 
\overline{H_{\alpha_1 \alpha_2} H_{\alpha_a \alpha_b}}\;\;\overline{H_{\alpha_2 \alpha_3} H_{\alpha_d \alpha_a}} \\
\times \;\;\overline{H_{\alpha_3 \alpha_4} H_{\alpha_b 
\alpha_c}}\;\;\overline{H_{\alpha_4 \alpha_1} H_{\alpha_c \alpha_d}} \;,\\
\overline{\lan ABCD\ran^m \lan ADBC\ran^m} = X_5 = \dis\f{1}{\l[d(f_m)\r]^2} \;\dis\sum_{\alpha_1, \alpha_2,\alpha_3,\alpha_4,\alpha_a,\alpha_b,\alpha_c,\alpha_d}\; 
\overline{H_{\alpha_1 \alpha_2} H_{\alpha_a \alpha_b}}\;\;\overline{H_{\alpha_2 \alpha_3} H_{\alpha_c \alpha_d}} \\
\times \;\;\overline{H_{\alpha_3 \alpha_4} H_{\alpha_d \alpha_a}}\;\;\overline{H_{\alpha_4 \alpha_1} H_{\alpha_b \alpha_c}} \;.
\earr \label{s44-5}
\ee 
Further simplification of these five terms (called $X_1$, $X_2$, $X_3$, $X_4$
and $X_5$ in Eq. (\ref{s44-3}) follow from the $SU(N)$ algebra given in
\cite{Ko-05,KM-strn,Butler,Butler1} and they involve several $SU(N)$
$U$-coefficients. However, formulas for these $U$-coefficients are not available in literature. For future reference we call the sum of the five terms as $X_{44}$,
\be
X_{44} = X_1 + X_2 + X_3 + X_4 + X_5
\label{s44-6}
\ee
Finally, the sixth term is
\be
\barr{l}
\overline{\lan ABCD\ran^m \lan ADCB\ran^m} = \dis\f{1}{\l[d(f_m)\r]^2} \;\dis\sum_{\alpha_1, \alpha_2,\alpha_3,\alpha_4,\alpha_a,\alpha_b,\alpha_c,\alpha_d}\; 
\overline{H_{\alpha_1 \alpha_2} H_{\alpha_a \alpha_b}}\;\;\overline{H_{\alpha_2 \alpha_3} H_{\alpha_d \alpha_a}} \\
\times \;\;\overline{H_{\alpha_3 \alpha_4} H_{\alpha_c \alpha_d}}\;\;\overline{H_{\alpha_4 \alpha_1} H_{\alpha_b \alpha_c}} \\
= \dis\f{1}{\l[d(f_m)\r]^2}\;\dis\sum_{\alpha_1, \alpha_2,\alpha_3,\alpha_4,\alpha_a,\alpha_b,\alpha_c,\alpha_d}\,\dis\sum_{\nu=1,\nu_2,\nu_3,\nu_4=0}^{m-k} \,\dis\sum_{\omega_{\nu_1},\omega_{\nu_2},\omega_{\nu_3},\omega_{\nu_4}}     \\
\Lambda^{\nu_1}(N,m,k)\, C^{\nu_1 , \omega_{\nu_1}}_{\alpha_1
\overline{\alpha_b}} C^{\nu_1, \omega_{\nu_1}}_{\alpha_a
\overline{\alpha_2}}\;\Lambda^{\nu_2}(N,m,k)\, C^{\nu_2 , \omega_{\nu_2}}_{\alpha_2 \overline{\alpha_a}}\;
C^{\nu_2 , \omega_{\nu_2}}_{\alpha_d \overline{\alpha_3}} \\
\Lambda^{\nu_3}(N,m,k)\, C^{\nu_3 , \omega_{\nu_3}}_{\alpha_3
\overline{\alpha_d}} C^{\nu_3, \omega_{\nu_3}}_{\alpha_c
\overline{\alpha_4}}\;\Lambda^{\nu_4}(N,m,k)\, C^{\nu_4 , \omega_{\nu_4}}_{\alpha_4 \overline{\alpha_c}}\;
C^{\nu_4 , \omega_{\nu_4}}_{\alpha_b \overline{\alpha_1}} \;. 
\earr \label{s44-7}
\ee
Now, simplifying the CG coefficients will give,
\be
\overline{\lan ABCD\ran^m \lan ADCB\ran^m} = \dis\f{1}{\l[d(f_m)\r]^2} \dis\sum_{\nu=0}^{m-k} \l[\Lambda^{\nu}(N,m,
k)\r]^4\,d(\nu)\;.
\label{s44-8}
\ee  
 
\renewcommand{\theequation}{C-\arabic{equation}}
\setcounter{equation}{0}   
\section*{APPENDIX C}

Formulas derived in Section \ref{sec4} contain finite $N$ corrections and they can be used to derive asymptotic limit formulas. These will provide a test, as often asymptotic formulas follow from a quite different formulation as given for example in \cite{MF,Br-81,SM}. To derive asymptotic formulas we will use the limit $N \rightarrow \infty$, $m \rightarrow \infty$,
$m/N \rightarrow 0$ and $k$ finite. Then we have the following approximations
\be
\barr{c}
\binom{N-p}{r} \stackrel{p/N \rightarrow 0}{\longrightarrow} \dis\f{N^r}{r!}\;,\;\;\;d(\nu) \stackrel{\nu /N \rightarrow 0}{\longrightarrow} \dis\f{N^{2\nu}}{(\nu !)^2}\;,\\
\Lambda^0(N,m,k) \rightarrow \binom{m}{k}\,\binom{N}{k}\;,\;\;\;
\Lambda^k(N,m,k) \rightarrow \binom{m-k}{k}\,\binom{N}{k}\;,\\
\Lambda^k(N,m,m-k) \rightarrow \binom{N}{m-k}\;,\;\;\;
\Lambda^0(N,m,m-k) \rightarrow \binom{m}{k}\,\binom{N}{m-k}\;.
\earr \label{asy1}
\ee
Using these, firstly we have 
\be
\Sigma_{2,0} = \binom{m}{k}\,\binom{N}{k}\;.
\label{asy2}
\ee
Using Eqs. (\ref{asy1}) and (\ref{asy2}) and the formulas given in Section \ref{sec4}, the following asymptotic limit formulas are obtained for $\sh_{PQ}$ with $(P,Q)=(1,1)$, $(3,1)$, $(2,2)$, $(5,1)$, $(4,2)$, $(3,3)$, $(7,1)$, $(6,2)$,
$(5,3)$ and $(4,4)$. These are,
\be
\barr{rcl}
\sh_{1,1} & = & \dis\f{\binom{m}{k}}{\binom{N}{k}^2}\;,\\
\sh_{3,1} & = & 3 \dis\f{\binom{m}{k}}{\binom{N}{k}^2} = 3\,\sh_{1,1} \;,\\
\sh_{2,2} & = & \dis\f{2}{\binom{N}{k}^2}\;.\\
\sh_{5,1} & = & 5 \dis\f{\binom{m}{k}}{\binom{N}{k}^2} \l[ 2 +
\,\dis\f{\binom{m-k}{k}}{\binom{m}{k}}\r] = (10+5q)\sh_{1,1} \;,\\
\sh_{4,2} & = & \dis\f{4}{\binom{N}{k}^2} \l[2 + \dis\f{\binom{m-k}{k}}{\binom{m}{k}}\r] = (4+2q)\sh_{2,2} \;,\\
\sh_{3,3} & = & 9 \dis\f{\binom{m}{k}}{\binom{N}{k}^2} + \dis\f{3}{\binom{m}{k}\,\binom{N}{k}^2} +\dis\f{3}{\binom{N}{k}^2}\,\l|U\r|^2 = 9 \sh_{1,1} + \dis\f{3}{\binom{m}{k}\,\binom{N}{k}^2} + O\l(\f{1}{\binom{N}{k}^4}\r) \;,\\
\sh_{7,1} & = & 7 \sh_{1,1} \;\l\{5 + 6 q + 3 q^2 +
\dis\f{\binom{m-2k}{k}}{\binom{m}{k}}(q) \r\} \simeq 7 \sh_{1,1} \;\l[5 + 6 q + 3 q^2 +q^3\r] \;,\\
\sh_{6,2} & = & \dis\f{1}{\binom{N}{k}^2} \l[30 + 36 q + 6 q^2 + 12U_1 +6U_2\r] \\
& = & \sh_{2,2} \l[15+18q+3q^2+9q^3\r] + O\l(\f{1}{\binom{N}{k}^4}\r) \;,\\
\sh_{5,3} & = & 15(2+q) \sh_{1,1} +5(1+q)\l[\sh_{3,3}-9\sh_{1,1}\r] +X_{53} \\
& = & 15\l\{(2+q)\sh_{1,1} +  \dis\f{(1+q)}{\binom{m}{k}\,\binom{N}{k}^2} \r\} + O\l(\f{1}{\binom{N}{k}^4}\r) \;,\\
\sh_{4,4} & = & 4(q+2)^2\,\sh_{2,2} + \dis\f{4}{\binom{m}{k}^2\,\binom{N}{k}^2} + X_{44} \\
& = & 4(q+2)^2\,\sh_{2,2} + \dis\f{4}{\binom{m}{k}^2\,\binom{N}{k}^2} + O\l(\f{1}{\binom{N}{k}^4}\r) \;;\\
q & = & \dis\f{\binom{m-k}{k}}{\binom{m}{k}}\;. 
\earr \label{asy4}
\ee
In the above equations, the following approximations (i)-(iv) are adopted. (i) $\l|U\r|^2$ in $\sh_{3,3}$ is the $U$-coefficient appearing in Eq. (\ref{fl-43a}) and it is expected to give negligible contribution to $\sh_{3,3}$. More importantly, the GUE formula (i.e. for $m=k$) for $\sh_{3,3}$ that can be derived easily shows that $|U|^2 \sim \binom{N}{k}^{-2}$ and this gives the final formula in Eq. (\ref{asy4}). (ii) In $\sh_{7,1}$, we used for the last term the approximation established in \cite{qMK-1}. 
(iii) Going further, $U_1$ and $U_2$ in $\sh_{6,2}$ are the terms with $U$-coefficients in $A2$ and $A3$ in Eq. (\ref{fl-58}). The GUE formulas and EGUE($k$) formulas assuming $\binom{m-k}{k}/\binom{m}{k} =1$ as given in \cite{Br-81} indicate the plausible result $6U_1 + 3U_2 = 9q^{3} + O\l(\f{1}{\binom{N}{k}^2}\r)$. 
(iv) From the GUE formulas it is plausible that $X_{53}$ and $X_{44}$ introduced in Appendix B will be of the order of $1/\binom{N}{k}^4$ and this is used in Eq. (\ref{asy4}) for $\sh_{5,3}$ and $\sh_{4,4}$. Finally, 
let us add that the diagrammatic method developed in \cite{SM} may hopefully give, in the near future, exact asymptotic limit formulas for $\sh_{6,2}$, $\sh_{5,3}$ and $\sh_{4,4}$ and for the last term in $\sh_{3,3}$.

\ed